\documentclass[]{pasj01}
\draft

\begin{document} 
\Received{}
\Accepted{}

\title{CO Multi-line Imaging of Nearby Galaxies (COMING):
VI. Radial variations in star formation efficiency}

 \author{%
   Kazuyuki \textsc{Muraoka}\altaffilmark{1},
   Kazuo \textsc{Sorai}\altaffilmark{2,3,4,5,6},
   Yusuke \textsc{Miyamoto}\altaffilmark{7},
   Moe \textsc{Yoda}\altaffilmark{8},
   Kana \textsc{Morokuma-matsui}\altaffilmark{9},
   Masato I.N. \textsc{Kobayashi}\altaffilmark{10},
   Mayu \textsc{Kuroda}\altaffilmark{1},
   Hiroyuki \textsc{Kaneko}\altaffilmark{11},
   Nario \textsc{Kuno}\altaffilmark{4,5},
   Tsutomu T. \textsc{Takeuchi}\altaffilmark{8},
   Hiroyuki \textsc{Nakanishi}\altaffilmark{12},
   Yoshimasa \textsc{Watanabe}\altaffilmark{4,5},
   Takahiro \textsc{Tanaka}\altaffilmark{4},
   Atsushi \textsc{Yasuda}\altaffilmark{4},
   Yoshiyuki \textsc{Yajima}\altaffilmark{3},
   Shugo \textsc{Shibata}\altaffilmark{3},
   Dragan \textsc{Salak}\altaffilmark{13},
   Daniel \textsc{Espada}\altaffilmark{7,14},
   Naoko \textsc{Matsumoto}\altaffilmark{7,15},
   Yuto \textsc{Noma}\altaffilmark{13},
   Shoichiro \textsc{Kita}\altaffilmark{4},
   Ryusei \textsc{Komatsuzaki}\altaffilmark{4},
   Ayumi \textsc{Kajikawa}\altaffilmark{6},
   Yu \textsc{Yashima}\altaffilmark{6},
   Hsi-An \textsc{Pan}\altaffilmark{16},
   Nagisa \textsc{Oi}\altaffilmark{17},
   Masumichi \textsc{Seta}\altaffilmark{13},
   and 
   Naomasa \textsc{Nakai}\altaffilmark{13}
}

 \altaffiltext{1}{Department of Physical Science, Osaka Prefecture University, Gakuen 1-1, Sakai, Osaka 599-8531, Japan}
 \email{kmuraoka@p.s.osakafu-u.ac.jp}
 \altaffiltext{2}{Department of Physics, Faculty of Science, Hokkaido University, Kita 10, Nishi 8, Kita-ku, Sapporo, Hokkaido 060-0810, Japan}
 \altaffiltext{3}{Department of Cosmosciences, Graduate School of Science, Hokkaido University, Kita 10, Nishi 8, Kita-ku, Sapporo, Hokkaido 060-0810, Japan}
 \altaffiltext{4}{Graduate School of Pure and Applied Sciences, University of Tsukuba, 1-1-1 Tennodai, Tsukuba, Ibaraki, 305-8571, Japan}
 \altaffiltext{5}{Tomonaga Center for the History of the Universe, University of Tsukuba, 1-1-1 Tennodai, Tsukuba, Ibaraki 305-8571, Japan}
 \altaffiltext{6}{Department of Physics, School of Science, Hokkaido University, Kita 10, Nishi 8, Kita-ku, Sapporo, Hokkaido 060-0810, Japan}
 \altaffiltext{7}{National Astronomical Observatory of Japan, 2-21-1 Osawa, Mitaka, Tokyo 181-8588, Japan}
 \altaffiltext{8}{Division of Particle and Astrophysical Science, Nagoya University, Furo-cho, Chikusa-ku, Nagoya, Aichi 464-8602, Japan}
 \altaffiltext{9}{Institute of Space and Astronautical Science, Japan Aerospace Exploration Agency, 3-1-1 Yoshinodai, Chuo-ku, Sagamihara, Kanagawa 252-5210, Japan}
 \altaffiltext{10}{Department of Earth and Space Science, Graduate School of Science, Osaka University, 1-1 Machikaneyama-cho, Toyonaka, Osaka 560-0043, Japan}
 \altaffiltext{11}{Nobeyama Radio Observatory, Minamimaki, Minamisaku, Nagano 384-1305, Japan}
 \altaffiltext{12}{Graduate School of Science and Engineering, Kagoshima University, 1-21-35 Korimoto, Kagoshima, Kagoshima 890-0065, Japan}
 \altaffiltext{13}{Department of Physics, School of Science and Technology, Kwansei Gakuin University, Gakuen 2-1, Sanda, Hyogo 669-1337, Japan}
 \altaffiltext{14}{The Graduate University for Advanced Studies (SOKENDAI), 2-21-1 Osawa, Mitaka, Tokyo, 181-0015, Japan}
 \altaffiltext{15}{The Research Institute for Time Studies, Yamaguchi University, Yoshida 1677-1, Yamaguchi, Yamaguchi 753-8511, Japan}
 \altaffiltext{16}{Academia Sinica, Institute of Astronomy and Astrophysics, No.1, Sec. 4, Roosevelt Rd, Taipei 10617, Taiwan}
 \altaffiltext{17}{Tokyo University of Science, Faculty of Science Division II, Liberal Arts, 1-3 Kagurazaka, Shinjuku-ku, Tokyo 162-8601, Japan}


\maketitle

\begin{abstract}
We examined radial variations in molecular-gas based star formation efficiency (SFE), which is defined as star formation rate per unit molecular gas mass,
for 80 galaxies selected from the CO Multi-line Imaging of Nearby Galaxies project (\cite{sorai2019}).
The radial variations in SFE for individual galaxies are typically a factor of 2 -- 3,
which suggests that SFE is nearly constant along galactocentric radius.
We found the averaged SFE in 80 galaxies of $(1.69 \pm 1.1) \times 10^{-9}$ yr$^{-1}$, which is consistent with \citet{leroy2008}
if we consider the contribution of helium to the molecular gas mass evaluation and the difference in the assumed initial mass function between two studies.
We compared SFE among different morphological (i.e., SA, SAB, and SB) types,
and found that SFE within the inner radii ($r/r_{25} < 0.3$, where $r_{25}$ is $B$-band isophotal radius at 25 mag arcsec$^{-2}$) of SB galaxies is slightly higher than that of SA and SAB galaxies.
This trend can be partly explained by the dependence of SFE on global stellar mass, which probably relates to the CO-to-H$_2$ conversion factor through the metallicity.
For two representative SB galaxies in our sample, NGC~3367 and NGC~7479, the ellipse of $r/r_{25}$ = 0.3 seems to cover not only the central region but also the inner part of the disk, mainly the bar.
These two galaxies show higher SFE in the bar than in spiral arms.
However, we found an opposite trend in NGC~4303; SFE is lower in the bar than in spiral arms, which is consistent with earlier studies (e.g., \cite{momose2010}).
These results suggest diversity of star formation activities in the bar.
\end{abstract}

\section{Introduction}

Star formation is one of the most fundamental processes in the evolution of galaxies.
Stars are formed by the contraction of the molecular interstellar medium,
which would be initiated by gravitational instability and/or cloud-cloud collision (e.g., \cite{mckee2007}).
Thus the observational study of the physical relation between molecular gas and star formation is indispensable to understand the evolution of galaxies.

The activity of star formation in galaxies is quantified as star formation rate (SFR).
Although the surface density of SFR is related to that of molecular gas via a power law (e.g.,\cite{kennicutt1998}),
its scatter is not so small; at a given surface gas density, the scatter of SFR spreads up to $\sim$ 1 dex even for nearby disk galaxies (e.g., \cite{daddi2010}).
In order to understand the physical origin of such a scatter, star formation efficiency (SFE), which is defined as SFR per unit molecular gas mass, is often investigated.
Many earlier studies pointed out that SFE differs among galaxies and even within a galaxy.
For example, \citet{young1996} reported that galaxies in disturbed environments (i.e., strongly interacting systems) show about four times higher SFE than in isolated spiral galaxies.
\citet{Lisenfeld2011} confirmed that SFE is a factor of 5 higher for a strongly perturbed sample and about 2 higher for a weakly perturbed sample compared to isolated galaxies.
For an individual galaxy, \citet{muraoka2007} found that SFE is four times higher in the central region than in the disk of the barred spiral galaxy M~83.
In addition, some earlier studies toward nearby galaxies reported that SFE is lower in bars than in spiral arms (e.g., \cite{momose2010}, \cite{watanabe2011}, \cite{hirota2014}, \cite{yajima2019}).

An extensive study of spatially-resolved SFE in galaxies has been conducted by \citet{leroy2008}.
They measured SFE at 800 pc resolution in 23 nearby galaxies using HERA CO-Line Extragalactic Survey (HERACLES) data,
and found that SFE based on H$_2$ is nearly constant at $(5.25 \pm 2.5) \times 10^{-10}$ yr$^{-1}$ as a function of various variables;
i.e., galactocentric radius, stellar mass surface density, gas surface density, free fall and orbital timescales, midplane gas pressure, stability of the gas disk to collapse,
the ability of perturbations to grow despite shear, and the ability of a cold phase to form, in H$_2$-dominated inner parts of spiral galaxies.
A similar trend is reported by \citet{bigiel2011}; they found a roughly constant molecular-gas depletion time,
which is the inverse of SFE, of $\sim$ 2.35 Gyr (corresponding to SFE of $4.26 \times 10^{-10}$ yr$^{-1}$) at 1 kpc resolution in HERACLES sample.
Recently, \citet{utomo2017} investigated the variation in molecular-gas depletion time on kpc-scales in 52 galaxies based on EDGE-CALIFA survey.
They found that 14 galaxies show a shorter depletion time of $\sim$ 1 Gyr (i.e., SFE of $10^{-9}$ yr$^{-1}$) in the center relative to that in the disk, $\sim$ 2.4 Gyr (SFE of $4.2 \times 10^{-10}$ yr$^{-1}$).
They also found that the central increase in SFE is correlated with a central increase in the stellar surface density,
suggesting that a higher SFE is associated with molecular gas compression by the stellar gravitational potential.

Currently, 147 CO($J=1-0$) maps of nearby galaxies, which are obtained by CO Multi-line Imaging of Nearby Galaxies (COMING) project (\cite{sorai2019}), have been available.
COMING is a project to map $J=1-0$ emission of $^{12}$CO, $^{13}$CO, and C$^{18}$O molecules toward a 70\% area of the optical disks of nearby galaxies
using the FOur-beam REceiver System on the 45 m Telescope (FOREST: \cite{minamidani2016}) at Nobeyama Radio Observatory (NRO).
Utilizing the initial CO data of COMING, the quantitative relationship between molecular gas and star formation has been examined for individual galaxies;
NGC~2903 \citep{muraoka2016}, NGC~2976 \citep{hatakeyama2017}, and NGC~4303 \citep{yajima2019}.
In particular, a positive correlation between SFE and molecular gas density is reported (\cite{muraoka2016}, \cite{yajima2019}).

We extend the study of the relation between molecular gas and star formation to many of COMING sample.
In particular, we focus on the difference in SFE both among galaxies and within each galaxy.
As a first step, we investigate radial variations in SFE of COMING sample.
Since COMING is the largest CO($J=1-0$) imaging survey of nearby galaxies, we can obtain the $general$ trend of variations in SFE with less bias.

The structure of this paper is as follows.
In section 2, we describe the selection of target galaxies and the data utilized to calculate SFE.
Then, we explain the derivation of radial profiles of SFE in subsection 3.1,
and show the results and their comparison to earlier studies in subsection 3.2.
In section 4, we firstly examine the dependence of SFE on the global stellar mass.
In addition, we investigate the difference in SFE among different morphological types of galaxies, i.e., SA, SAB, and SB galaxies.
Finally, we provide some implications regarding star formation in the inner part of disk galaxies; in particular, star formation in the bar.

\section{Sample selection and data}

In this section, we describe selection criteria of sample galaxies from COMING and summarize the data utilized to calculate SFE.

\subsection{Sample selection}

In order to select appropriate galaxies for examining radial variations in SFE from COMING sample,
we set three criteria as follows.
(1) The inclination ($i$) is less than 75$^{\circ}$.
(2) The obvious interaction of galaxies is not observed.
(3) SFR can be calculated by a combination of GALEX FUV \citep{gil2007} and WISE 22 $\mu$m \citep{wright2010} data.
The first criterion is set because it is difficult to accurately derive azimuthally-averaged radial profiles of molecular gas mass and SFR toward highly-inclined galaxies.
The second is set to avoid enhancement or suppression of star formation by the interaction of galaxies.
The third is to obtain homogeneous SFR maps among selected galaxies, and to discuss the difference in SFE without systematic errors caused by different calibration methods of SFR.
Note that we did not employ MIPS 24 $\mu$m data as a tracer of SFR because the number of corresponding 24 $\mu$m maps for COMING sample is smaller than that of WISE 22 $\mu$m maps
and because some 24 $\mu$m maps suffer from saturation in bright regions.

Based on the three criteria, we finally selected 80 galaxies from COMING sample in total; 30 SA, 33 SAB, and 17 SB galaxies.
Parameters for 80 selected galaxies, i.e., galaxy name, $B$-band isophotal radius at 25 mag arcsec$^{-2}$ ($r_{25}$), inclination $i$, position angle (P.A.), and global stellar mass ($M_{\ast}$) are summarized in table~1.
Figure~1 shows histograms of $M_{\ast}$ for three galaxy types. Compared to SA and SAB galaxies, selected SB galaxies are less massive.
This trend also appears in the original COMING sample \citep{sorai2019}.

\subsection{Data}

We produce maps of the H$_2$ mass surface density $\Sigma_{\rm H_2}$ and the SFR surface density $\Sigma_{\rm SFR}$ to obtain radial profiles of SFE for selected galaxies.

\subsubsection{H$_2$ mass}

$\Sigma_{\rm H_2}$ is calculated from $^{12}$CO($J=1-0$) integrated intensities obtained by COMING at an angular resolution of 17$''$,
which corresponds to the spatial resolution of 0.3 -- 3 kpc (typically $\sim$ 1.5 kpc) for our sample.
A detailed description of the COMING observations, data processing, and CO maps are presented in the COMING overview paper \citep{sorai2019}.
We uniformly assume the standard CO-to-H$_2$ conversion factor $X_{\rm CO}$ = $2.0 \times 10^{20}$ cm$^{-2}$ (K km s$^{-1}$)$^{-1}$ \citep{bolatto2013}
for all the selected galaxies. In addition, no spatial variation in $X_{\rm CO}$ is assumed within each galaxy.
Using a factor of cos($i$), $\Sigma_{\rm H_2}$ are corrected (i.e., deprojected) for each galaxy.
Note that the contribution of helium mass (a factor of 1.36) is $not$ included in our $\Sigma_{\rm H_2}$.

\subsubsection{Star formation rate}

$\Sigma_{\rm SFR}$ is calculated from a linear combination of GALEX FUV and WISE 22 $\mu$m intensities using the following formula \citep{casasola2017}:
\begin{eqnarray}
\left[ \frac{{\Sigma_{\rm SFR}}}{M_{\odot} \,\, {\rm yr}^{-1} \,\, {\rm kpc}^{-2}} \right] &=& \left[ 3.2 \times 10^{-3} \left( \frac{I_{22}}{{\rm MJy \,\, sr}^{-1}} \right) + 8.1 \times 10^{-2} \left( \frac{I_{\rm FUV}}{{\rm MJy \,\, sr}^{-1}} \right) \right] \nonumber \\
&& \times {\rm cos} \, i \times 1.59
\end{eqnarray}
where $I_{22}$ and $I_{\rm FUV}$ are 22 $\mu$m and FUV intensities, respectively. 
A factor of 1.59 converts the obtained $\Sigma_{\rm SFR}$ to those shown by \citet{kennicutt1998} assuming the truncated \citet{salpeter1955} initial mass function (IMF).

For FUV data, we firstly estimated the Galactic dust extinction at 1538.6 \AA, which is the effective wavelength of FUV band,
based on the Galactic $V$-band extinction A$_V$ (\cite{schlafly2011}) and \citet{cardelli1989} model, and corrected the dust extinction.
Then, we convolved the FUV images at an angular resolution of $4''.2$ to $17''$ to match COMING data.
Finally, we converted the unit of original FUV images, count pixel second (CPS), into MJy sr$^{-1}$.

For WISE 22 $\mu$m data, we individually determined the background level as an average of the intensity value at the blank sky, and subtracted it.
Then, we converted the unit of original 22 $\mu$m images, DN, into MJy sr$^{-1}$.
Note that we did not convolve 22 $\mu$m images because the effective angular resolution of 22 $\mu$m Atlas images is 16".8 \citep{cutri2012},
which is almost the same as that of COMING data.

We finally obtained the SFR map with an angular resolution of 17$''$.
The details of treating FUV and 22 $\mu$m data and calculating $\Sigma_{\rm SFR}$ are described in Yoda et al. (in preparation).

\section{Radial profiles of SFE}

\subsection{Derivation of radial profiles}

Using the maps of $\Sigma_{\rm H_2}$ and $\Sigma_{\rm SFR}$, we obtain radial profiles of SFE as described below.
Firstly, both maps with a grid size of 6$''$ per pixel are regridded to 1$''$.2 per pixel.
Secondly, we determine a tilted ring of a galaxy based on its position angle and inclination.
Then, we set 6$''$ wide tilted rings as bins of a radial profile, and calculate azimuthally-averaged $\Sigma_{\rm H_2}$ and $\Sigma_{\rm SFR}$ in each bin.
Finally, SFE in each bin is obtained by the division of $\Sigma_{\rm SFR}$/$\Sigma_{\rm H_2}$.
We normalize the unit of galactocentric radius by $r_{25}$, and thus we show SFE as a function of $r/r_{25}$.
For barred spiral (SAB and SB) galaxies, the bar-end roughly corresponds to $r/r_{25}$ of 0.2 -- 0.3.

We treated the uncertainties in $\Sigma_{\rm H_2}$, $\Sigma_{\rm SFR}$, and SFE for each bin in the manner of \citet{leroy2008}.
They took the uncertainty $\sigma$ in a quantity averaged over a tilted ring to be
\begin{eqnarray}
\sigma = \frac{\sigma_{\rm rms}}{\sqrt{N_{\rm pix,ring}/N_{\rm pix,beam}}}
\end{eqnarray}
where $\sigma_{\rm rms}$ is the r.m.s.\ scatter within the tilted ring, $N_{\rm pix,ring}$ is the number of pixels included in the ring, and $N_{\rm pix,beam}$ is the number of pixels included in the observing beam size.
However, the obtained $\sigma$ for $\Sigma_{\rm SFR}$ seemed too small (typically $<$ 1\%), and thus we considered the systematic uncertainties as $\sigma$ for $\Sigma_{\rm SFR}$.
According to \citet{gil2007}, the error in the FUV zero-point calibration is estimated to be 0.15 mag, which yields the uncertainty of $\sim$ 15\%.
In addition, \citet{jarrett2011} reported that the rms scatter about the zero level is 5.7\% for WISE 22 $\mu$m band.
Considering that WISE 22 $\mu$m intensity is dominant in $\Sigma_{\rm SFR}$ (typically $>$ 80\%) compared to FUV intensity,
we estimated that the systematic uncertainty of $\Sigma_{\rm SFR}$ is typically 8\%.

We calculated $\sigma$ of $\Sigma_{\rm H_2}$ and the systematic uncertainty of $\Sigma_{\rm SFR}$ in each bin,
and finally derived the uncertainty in SFE from the error propagation.
Figure~2 shows an example of radial profiles of $\Sigma_{\rm H_2}$, $\Sigma_{\rm SFR}$, and SFE for NGC~4303.
We display each measurement (data point) in bins where the calculated SFE is more than 3 $\sigma$.

\subsection{Results}

Figures~3, 4, and 5 show the radial profiles of SFE for individual SA, SAB, and SB galaxies, respectively.
Most of galaxies do not show drastic variations in SFE except for three barred spiral galaxies (NGC~1087, NGC~2268, and NGC~4579);
the radial variation in SFE is less than a factor of 5 (typically a factor of 2 -- 3).
This suggests that molecular-gas based SFE is nearly constant along galactocentric radius (see also figures~6 and 7), which is consistent with the finding by \citet{leroy2008}.

We found the averaged SFE in 80 galaxies of $(1.69 \pm 1.1) \times 10^{-9}$ yr$^{-1}$, which seems significantly higher than that of $(5.25 \pm 2.5) \times 10^{-10}$ yr$^{-1}$ reported by \citet{leroy2008}.
This is because there are two differences in the calculation of SFE between the two studies;
firstly, the contribution of helium mass (a factor of 1.36) is not included in our $\Sigma_{\rm H_2}$.
In addition, our $\Sigma_{\rm SFR}$ is systematically higher than that in \citet{leroy2008} by a factor of 1.59 because they assumed a \citet{kroupa2001} IMF.
Therefore, our SFE must be divided by 2.16 when it is compared with the SFE in \citet{leroy2008}.
This yields the averaged SFE of $(7.82 \pm 5.0) \times 10^{-10}$ yr$^{-1}$ in this study, which is consistent with that of \citet{leroy2008} within the error ranges.

In figure~4, NGC~3310 shows extremely high SFE ($> 10^{-8}$ yr$^{-1}$).
This is partly because of an intense starburst likely caused by a recent minor interaction (\cite{miralles2014}), which yields an elevated SFR.
In addition, NGC~3310 shows a moderately low metallicity (0.2 -- 0.4 $Z_{\odot}$; \cite{pastoriza1993}),
which suggests higher $X_{\rm CO}$ than the standard value and thus the underestimation of $\Sigma_{\rm H_2}$.
For these reasons, such an extremely high SFE is observed in NGC~3310.

\section{Discussion}

In this section, we examine the general trend of radial variations in SFE for each galaxy type, i.e., SA, SAB, and SB galaxies.
Then, we evaluate its difference among the three types.
In particular, we discuss an unexpected trend of SFE; i.e., higher SFE within the inner radii of SB galaxies compared to SA and SAB galaxies.

\subsection{General trend of SFE and its dependence on global stellar mass}

Figure~6 shows the compiled radial profiles of SFE for each galaxy type.
As described in subsection 3.2, SFE seems nearly constant along galactocentric radius.
In order to evaluate how radial profiles are affected by the difference in spatial resolution among our sample,
we examined radial profiles of SFE at a common spatial resolution of 3 kpc as shown in figure~7.
Radial profiles shown in figures 6 and 7 seem closely similar to each other,
and thus we concluded that the difference in spatial resolution does not affect the subsequent discussion.

We found a possibility that SFE in the inner radii of SB galaxies is higher than that of SA and SAB galaxies.
To confirm this trend clearly, we examine averaged SFE over a 0.1 $r/r_{25}$ wide for every galaxy types as shown in figure~8.
In the inner radii ($r/r_{25} < 0.3$), SFE of SB galaxies ($> 2.0 \times 10^{-9}$ yr$^{-1}$) is slightly higher than that of SA and SAB galaxies ($\sim 1.4 \times 10^{-9}$ yr$^{-1}$).
On the other hand, in the outer radii ($0.3 < r/r_{25} < 0.5$), all the three types show similar SFEs of (1.3 -- 1.7) $\times 10^{-9}$ yr$^{-1}$.
Note that the number of data points for SB galaxies in figure 6 is smaller than SA and SB galaxies due to the difference in sample size.
This might cause the averaged radial profile for SB galaxies deviating from other types.
In particular, such a deviation seems evident in the outermost radii ($r/r_{25} > 0.5$);
SFE of SB galaxies becomes slightly higher than SA and SAB galaxies again at the radii, whereas this trend seems doubtful because there are few data points as shown in figure~6.
Such an effect due to the difference in sample size is also discussed in subsection 4.2.

Here, we investigate why SB galaxies show higher SFE within the inner radii ($r/r_{25} < 0.3$) compared to SA and SAB galaxies.
Firstly, we examine the dependence of SFE on global stellar mass $M_{\ast}$ because selected SB galaxies in this study are less massive compared to SA and SAB galaxies as described in subsection 2.1.
Figure~9 shows the comparison between SFE within the inner radii and $M_{\ast}$ in our sample.
A trend that SFE decreases with the increase in $M_{\ast}$ can be seen.
We obtain the Spearman rank correlation coefficient of $-0.52$ for this log SFE -- log $M_{\ast}$ relation, suggesting a significant negative correlation.
Such a negative correlation between SFE and $M_{\ast}$ in the stellar mass range of $10^9$ to $10^{11.5}$ $M_{\odot}$ has been reported in earlier studies.
For example, \citet{saintonge2011} and \citet{huang2014} found that molecular-gas depletion time becomes longer (i.e, SFE becomes lower) with the increase in stellar mass based on COLD GASS.
In addition, \citet{leroy2013} and \citet{boselli2014} reported similar relationships based on IRAM HERACLES and Herschel Reference Survey, respectively.
This correlation is presumably due to the relationship between stellar mass and metallicity,
i.e., so-called mass-metallicity relation (e.g., \cite{lequeux1979}, \cite{tremonti2004}, \cite{hughes2013}).
Both parameters can trace the integrated star formation history within a galaxy;
star formation firstly increases stellar mass, and then, heavy elements are ejected from evolved stars into interstellar space.
The increase in metallicity results in the decrease in $X_{\rm CO}$ (e.g., \cite{arimoto1996}).
Since we assume the standard $X_{\rm CO}$ for all the sample, 
we may overestimate $\Sigma_{\rm H_2}$ for galaxies with high metallicity, which eventually results in an underestimation of SFE.
In this case, a negative correlation between SFE and $M_{\ast}$ may appear.
In fact, the negative correlation between SFE and the metallicity is already found (e.g., \cite{dib2011}, \cite{leroy2013}).

Therefore, the obtained negative correlation between SFE and $M_{\ast}$ can be interpreted as an apparent variation in SFE due to the difference in $X_{\rm CO}$ among our sample.
Such an explanation is also reported by \citet{leroy2013}.
However, we cannot rule out another possibility that SFE truly varies depending on galaxy mass.
For example, \citet{saintonge2011} suggest that smaller galaxies have more bursty star formation histories due to minor starburst events,
and also suggest that quenching in high mass galaxies prevent the molecular gas from forming stars without destroying the gas.
To clarify whether the negative correlation between SFE and $M_{\ast}$ is attributed to the change in $X_{\rm CO}$ or that in true SFE,
an independent determination of $X_{\rm CO}$ (e.g., simultaneous solving the CO-to-H$_2$ conversion factor and dust-to-gas ratio proposed by \cite{sandstrom2013})
in each galaxy and its comparison with the observed SFE are necessary.

\subsection{Is SFE within the inner radii of SB galaxies intrinsically high?}

Since the dependence of SFE on $M_{\ast}$ has been confirmed, we try to cancel it and to evaluate the residual of SFE.
We examined the log SFE -- log $M_{\ast}$ relation by the ordinary least square power-law fit without uncertainties,
and obtained the expression of ${\rm log \,\, SFE_{fit}} = -0.424 \times ({\rm log} \,\, M_{\ast}) - 4.31$.
Then, we define the residual of SFE within the inner radii ($r/r_{25} < 0.3$) as 
\begin{eqnarray}
\Delta ({\rm log \,\, SFE}) = {\rm log \,\, SFE_{obs}} - {\rm log \,\, SFE_{fit}},
\end{eqnarray}
where SFE$_{\rm obs}$ is the observed SFE.
Figure~10 shows the comparison between $\Delta$(log SFE) and $M_{\ast}$. Most of galaxies are within the $\Delta$(log SFE) range of $\pm$ 0.5.
In order to make the distribution of $\Delta$(log SFE) more visible, we examine histograms of $\Delta$(log SFE) for each galaxy type as shown in figure~11.
The frequency distributions of SA and SAB galaxies seem quite similar to each other, 
whereas that for SB galaxies seems offset to the high-SFE side compared to SA and SAB galaxies.

To evaluate whether the frequency distributions of $\Delta$(log SFE) are different among SA, SAB, and SB galaxies, we examine two-sample Kolmogorov-Smirnov (K-S) tests.
The K-S test between SA and SAB galaxies gives a $p$ value of 0.51, and thus the hypothesis that the two samples originate from the same distribution is not rejected.
However, the K-S test between SAB and SB galaxies gives a $p$ value of 0.037
\footnote{The K-S test between SA and SB galaxies gives a $p$ value of 0.024.}.
This means the hypothesis that SAB and SB samples originate from the same distribution is rejected with 95\% confidence interval,
suggesting that SFE within the inner radii ($r/r_{25} < 0.3$) of SB galaxies is significantly higher than that of SA and SAB galaxies even though the dependence of SFE on $M_{\ast}$ is cancelled.

Generally, stellar bars in galaxies can efficiently transport a large amount of molecular gas toward central regions (e.g., \cite{matsuda1977}, \cite{simkin1980}, \cite{sakamoto1999}, \cite{haan2009}),
which invokes a nuclear star formation (e.g., \cite{jogee2005}, \cite{schinnerer2006}).
Since SB galaxies typically have strong bars, high-SFE nuclear star formation (i.e., starburst) likely occurs in SB galaxies.
However, the spatial scale of the nuclear star formation in nearby galaxies is typically in the range of a few $\times$ 100 pc to 1 kpc (e.g., \cite{kohno1999}, \cite{schinnerer2006}, \cite{muraoka2009}), which corresponds to $r/r_{25} \sim 0.1$ or less.
It seems that such a bursty star formation should occur even in the disk region to explain the high SFE toward $r/r_{25} \sim 0.3$.

Note that increasing sample galaxies could affect our results.
If only an SB galaxy with $\Delta$(log SFE) $= 0$ is added to our sample, 
the $p$ value of the K-S test between SAB and SB galaxies increases up to 0.058, which means that the null hypothesis is not rejected.
Therefore, it is important to increase the number of SB galaxies for a more robust understanding of the difference in SFE among galaxy types.

\subsection{High SFE in the bar of SB galaxies}

In order to investigate the physical origin of the high SFE within the inner radii ($r/r_{25} < 0.3$) further,
we examine SFE maps for two representative SB galaxies in our sample, NGC~3367 and NGC~7479, as shown in figure~12.
The ellipse of $r/r_{25}$ = 0.3 seems to cover not only the central region but also a part of the disk, mainly the bar.
For NGC~3367, we found that SFEs are (5 -- 8) $\times 10^{-9}$ yr$^{-1}$ in the central region,
(2 -- 4) $\times 10^{-9}$ yr$^{-1}$ at $r/r_{25}$ of 0.1 -- 0.3 (corresponding to the bar),
and (1 -- 2) $\times 10^{-9}$ yr$^{-1}$ at $r/r_{25} > 0.3$ (corresponding to spiral arms) with the typical uncertainty of $0.5 \times 10^{-9}$ yr$^{-1}$.
This indicates that SFE in the bar is lower than that in the central region, but higher than in spiral arms.
A similar trend is found in NGC~7479; SFEs are (3 -- 6) $\times 10^{-9}$ yr$^{-1}$ in the central region,
(3 -- 6) $\times 10^{-9}$ yr$^{-1}$ at the northern bar, (1 -- 2) $\times 10^{-9}$ yr$^{-1}$ at the southern bar,
and (0.5 -- 2) $\times 10^{-9}$ yr$^{-1}$ in spiral arms with the typical uncertainty of $0.3 \times 10^{-9}$ yr$^{-1}$.

At least some of SB galaxies in our sample show higher SFE in the bar than in spiral arms.
However, earlier studies reported an opposite trend in nearby barred spiral galaxies;
SFE is lower in the bar than in spiral arms (e.g., \cite{momose2010}, \cite{watanabe2011}, \cite{hirota2014}, \cite{yajima2019}).
Here, we compare our SFE in NGC~4303 with that reported by \citet{momose2010}.
They adopted the bar radius of 40$''$ in NGC~4303, which corresponds to $r/r_{25}$ of 0.2.
Thus we examine the averaged SFE separating NGC~4303 disk into the inner part ($r/r_{25} < 0.2$) and the outer part ($r/r_{25} > 0.2$).
We obtain the averaged SFE in the inner part of $(8.4 \pm 0.9) \times 10^{-10}$ yr$^{-1}$ and that in the outer part of $(1.2 \pm 0.1) \times 10^{-9}$ yr$^{-1}$ (see figure~2);
i.e., SFE in the bar is $\sim$ 1.4 times lower than that in spiral arms.
This trend is consistent with \citet{momose2010}, but opposite to NGC~3367 and NGC~7479.
Eventually, our results suggest that some barred spiral galaxies surely show lower SFE in the bar than in spiral arms as reported by earlier studies, whereas this trend is not always true for other barred spiral galaxies.
In other words, there might be diversity of star formation activities in the bar.

Some mechanism to suppress star formation in the bar has been already proposed.
Generally, strong shocks and shear induced by the non-circular motion in the bar suppress the formation of dense molecular gas and massive stars,
and thus SFE becomes lower in the bar compared to spiral arms (e.g., \cite{reynaud1998}).
In addition, recent numerical simulations suggest that star formation is enhanced by cloud-cloud collisions in moderate velocity (15 -- 40 km s$^{-1}$), which typically occur in spiral arms,
but is rather suppressed by those in high velocity ($>$ 60 km s$^{-1}$), which often occur in the bar (\cite{fujimoto2014}). This eventually causes lower SFE in the bar than in spiral arms.
In fact, the latest CO observations confirmed that the velocity dispersion of molecular gas, which would be an indicator of collision velocity of clouds, is larger in the bar than in spiral arms (e.g., \cite{maeda2018}, \cite{yajima2019}).

It is difficult to explain the physical origin of the high SFE in the bar, but a hint is provided by \citet{zhou2015}.
They reported that barred galaxies with intense star formation in bars tend to have active star formation also in their bulges and disks.
This suggests a possibility that a bar-driven process enhances SFE not only in the central region but also in the bar,
although the specific physical mechanism to enhance star formation in the bar is unclear yet.

Note that we have to consider why SFEs of SA and SAB galaxies are similar to each other in all the radii
although there is a distinct difference between the two types; existence or non-existence of the bar.
According to the explanation for the higher SFE in SB galaxies than in SAB galaxies as described above,
we likely observe higher SFE in SAB galaxies than in SA galaxies because SAB galaxies have bars.
In fact, earlier studies reported that the degree of molecular-gas concentration to the central kiloparsec is higher in barred galaxies than in non-barred galaxies (e.g., \cite{sakamoto1999}, \cite{kuno2007})
\footnote{In \citet{sakamoto1999} sample, 9 out of 10 barred galaxies are classified into SAB type, and 24 out of 28 barred galaxies are SAB type in \citet{kuno2007} sample.
Thus we can read their studies as the comparison between SA and SAB galaxies.}.
This suggests that intense star formation with high SFE occurs in the central regions of SAB galaxies as well as SB galaxies, whereas our results do not support such a simple scenario.

In order to solve such complications regarding the difference in SFE within the inner radii among SA, SAB, and SB galaxies,
formation processes of dense gas and massive stars in the bar should be investigated further.
In particular, it is essential to understand dynamical effects such as cloud-cloud collisions, shocks, and shear
because they play a key role in enhancement or suppression of star formation.
In addition, the measurement of fractional mass of the dense gas in molecular gas ($f_{\rm DG}$) is important.
\citet{torii2018} examined $f_{\rm DG}$ in the Milky Way using $^{12}$CO and C$^{18}$O molecules,
and found that $f_{\rm DG}$ in the Galactic arms is as high as $\sim$ 4 -- 5\%, while it becomes quite small at 0.1 -- 0.4\% in the Galactic bar and inter-arm regions.
Such a study of $f_{\rm DG}$ and its relation to dynamical effects can be conducted for external galaxies with high-sensitivity and high-angular resolution observations.
The latest interferometers, such as Atacama Large Millimeter/submillimeter Array and the Northern Extended Millimeter Array,
are certainly helpful for the further understanding of the star formation in the bar.

\section{Summary}

We examined radial variations in molecular-gas based SFE for 80 galaxies selected from COMING sample.
A summary of this work is as follows.

\begin{enumerate}
\item
We found that the radial variations in SFE for individual galaxies are typically a factor of 2 -- 3,
which suggests that SFE is nearly constant along galactocentric radius.

\item
We found the averaged SFE in 80 galaxies of $(1.69 \pm 1.1) \times 10^{-9}$ yr$^{-1}$, which is consistent with \citet{leroy2008}
if we consider the contribution of helium to the molecular gas mass evaluation and the difference in the assumed initial mass function between two studies.

\item
Within the inner radii ($r/r_{25} < 0.3$), SFE of SB galaxies ($> 2.0 \times 10^{-9}$ yr$^{-1}$) is significantly higher than that of SA and SAB galaxies ($\sim 1.4 \times 10^{-9}$ yr$^{-1}$).
This trend can be partly explained by the dependence of SFE on global stellar mass, which probably relates to $X_{\rm CO}$ through the metallicity.
According to the K-S test, however, we found that SFE within the inner radii of SB galaxies is still higher than that of SA and SAB galaxies even though the dependence of SFE on $M_{\ast}$ is cancelled.

\item
For two representative SB galaxies in our sample, NGC~3367 and NGC~7479, the ellipse of $r/r_{25}$ = 0.3 seems to cover not only the central region but also a part of the disk, mainly the bar.
These two galaxies show higher SFE in the bar than in spiral arms.
However, we found an opposite trend in NGC~4303; SFE is lower in the bar than in spiral arms, which is consistent with earlier studies (e.g., \cite{momose2010}).
These results suggest diversity of star formation activities in the bar.

\end{enumerate}

\vspace{0.5cm}
We thank the referee for invaluable comments, which significantly improved the manuscript.
We are indebted to the NRO staff for the commissioning and operation of the 45-m telescope and their continuous efforts to improve the performance of the instruments.
This work is based on observations at NRO, which is a branch of the National Astronomical Observatory of Japan, National Institutes of Natural Sciences.
This research has made use of the NASA/IPAC Extragalactic Database, which is operated by the Jet Propulsion Laboratory,
California Institute of Technology, under contract with the National Aeronautics and Space Administration.
This work was supported by JSPS KAKENHI (Grant Nos.\ 17K14251 and 18J00508).


\begin{figure}
  \begin{center}
    \includegraphics[width=8cm]{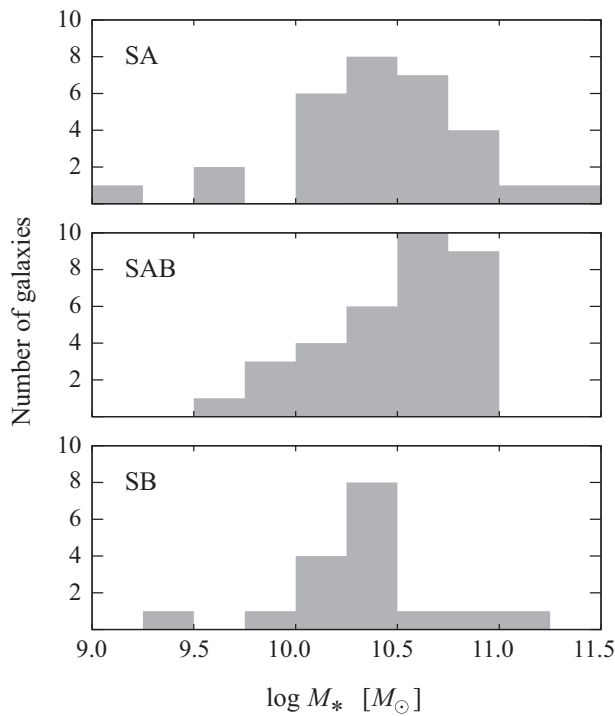}
  \end{center}
\caption{
Histograms of stellar mass $M_{\ast}$ for selected SA, SAB, and SB galaxies.
There are few SB galaxies with $M_{\ast}$ of more than $10^{10.5} M_{\odot}$.
}
\label{fig:fig1}
\end{figure}

\begin{figure}
  \begin{center}
    \includegraphics[width=8cm]{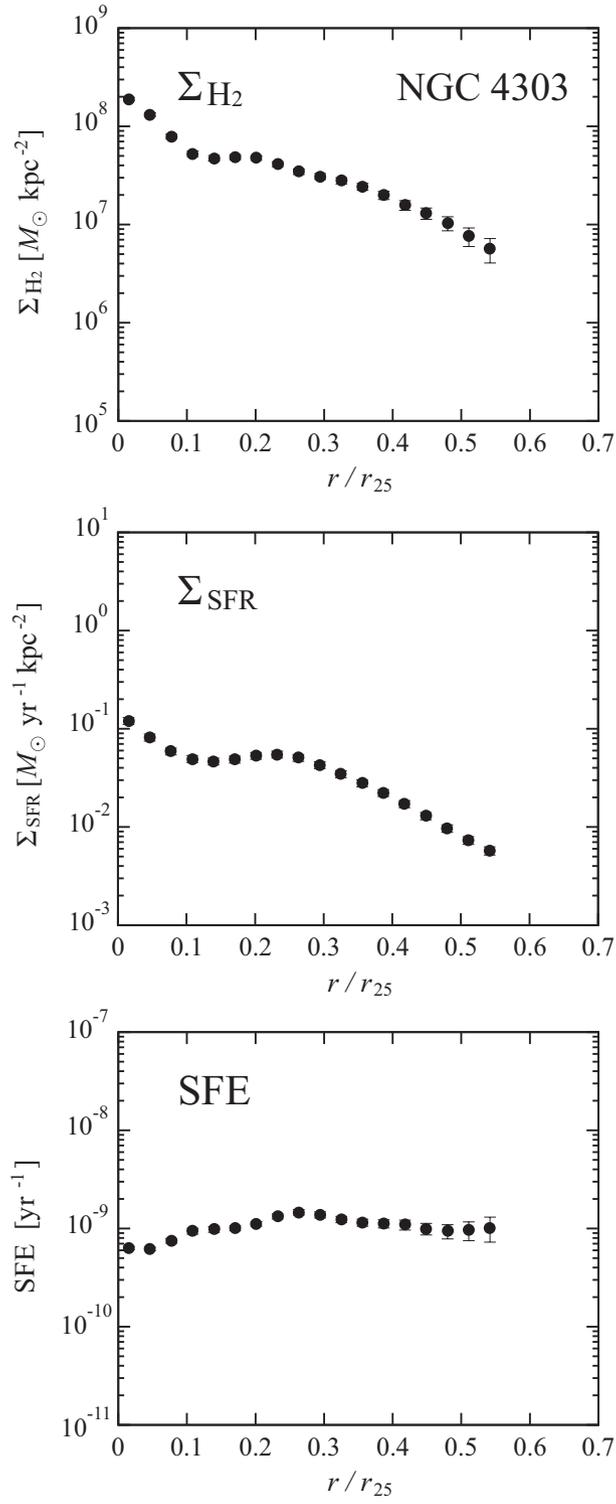}
  \end{center}
\caption{
An example of radial profiles of $\Sigma_{\rm H_2}$ (top), $\Sigma_{\rm SFR}$ (middle), and SFE (bottom) for NGC~4303.
}
\label{fig:fig2}
\end{figure}

\begin{figure}
  \begin{center}
    \includegraphics[width=17cm]{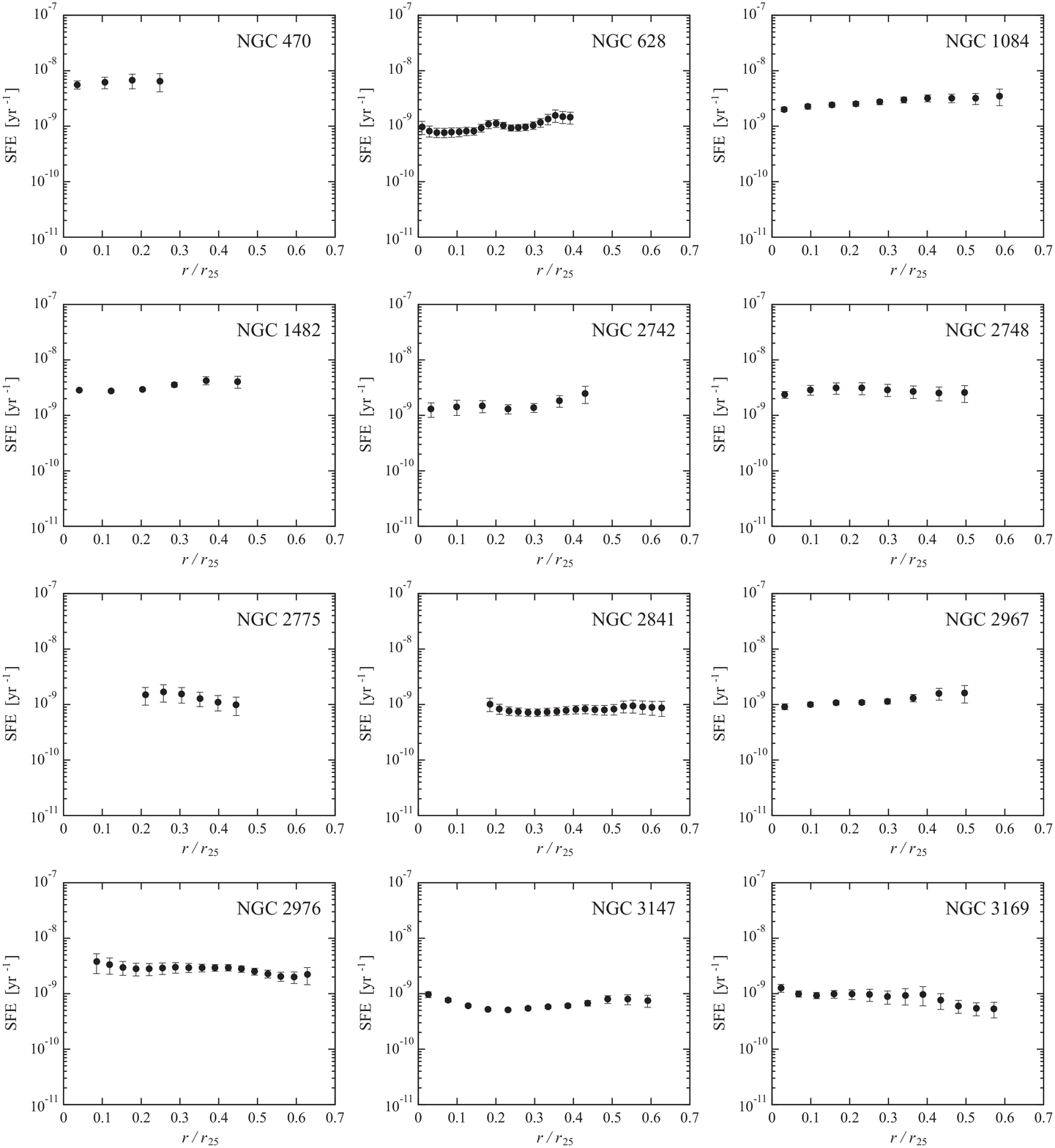}
  \end{center}
\caption{
SFE as a function of the galactocentric radius in the unit of $r/r_{25}$ for individual SA galaxies.
}
\label{fig:fig3a}
\end{figure}

\setcounter{figure}{2}
\begin{figure}
  \begin{center}
    \includegraphics[width=17cm]{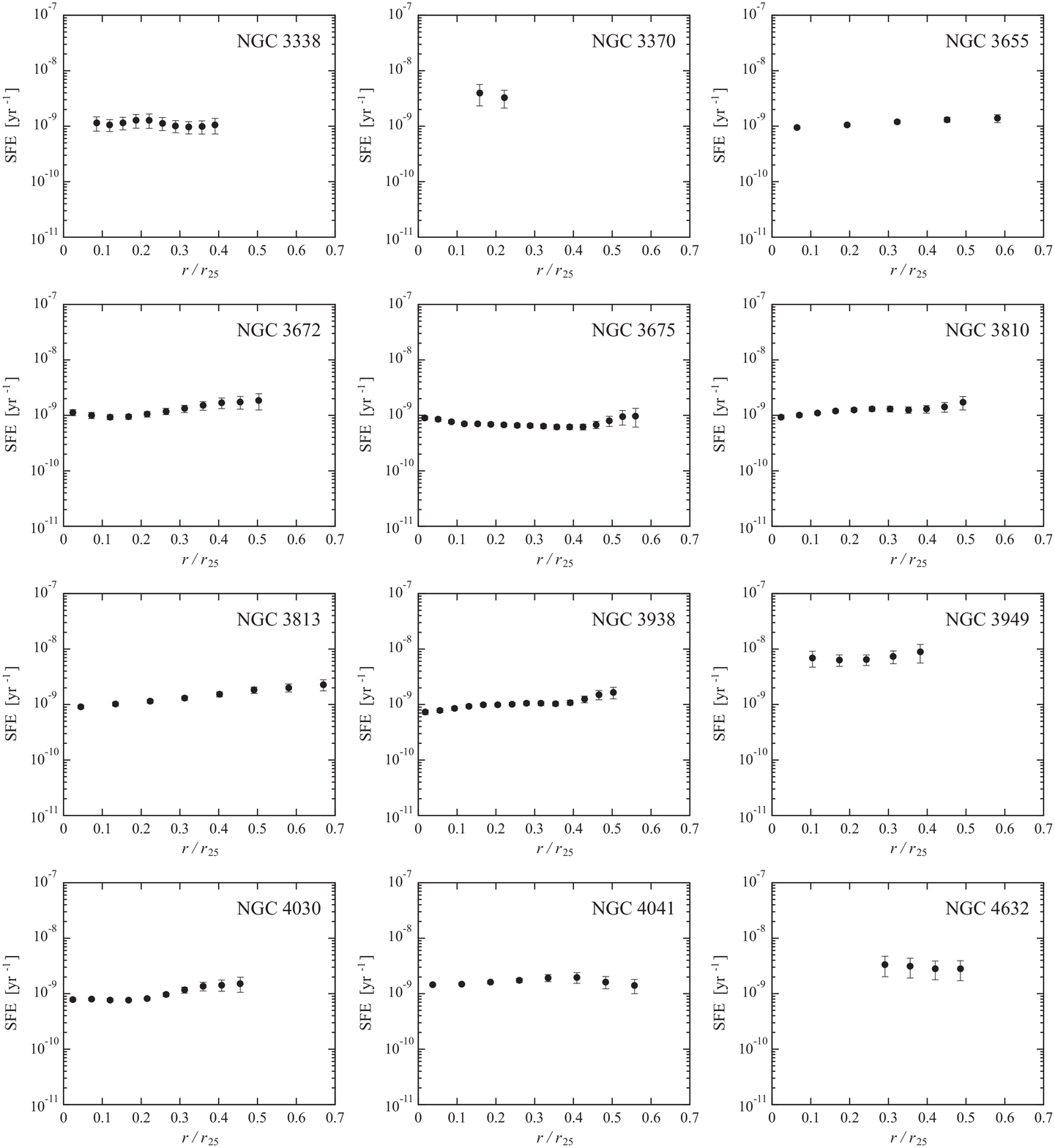}
  \end{center}
\caption{
(Continued)
}
\label{fig:fig3b}
\end{figure}

\setcounter{figure}{2}
\begin{figure}
  \begin{center}
    \includegraphics[width=17cm]{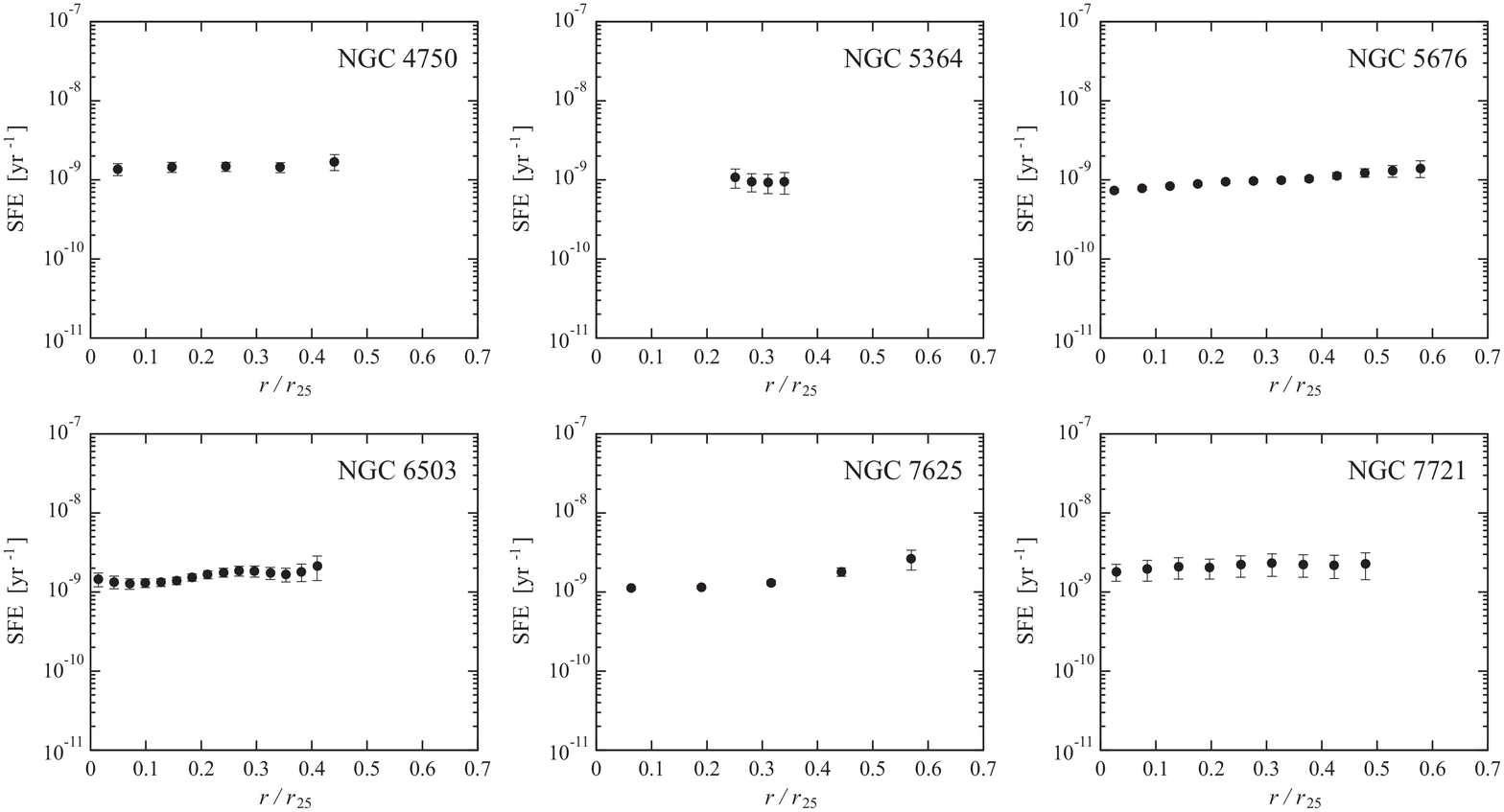}
  \end{center}
\caption{
(Continued)
}
\label{fig:fig3c}
\end{figure}

\begin{figure}
  \begin{center}
    \includegraphics[width=17cm]{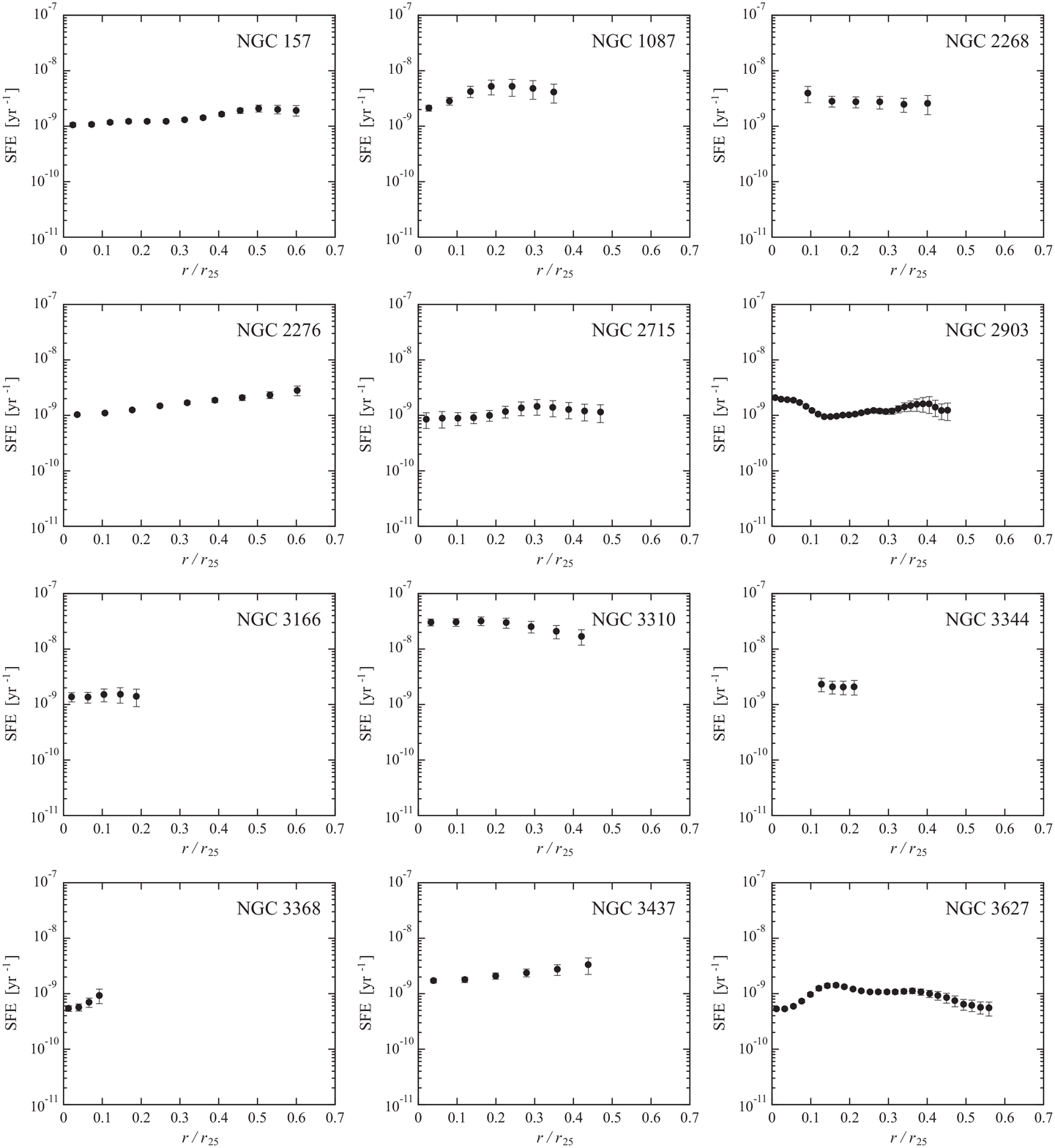}
  \end{center}
\caption{
SFE as a function of the galactocentric radius in the unit of $r/r_{25}$ for individual SAB galaxies.
}
\label{fig:fig4a}
\end{figure}

\setcounter{figure}{3}
\begin{figure}
  \begin{center}
    \includegraphics[width=17cm]{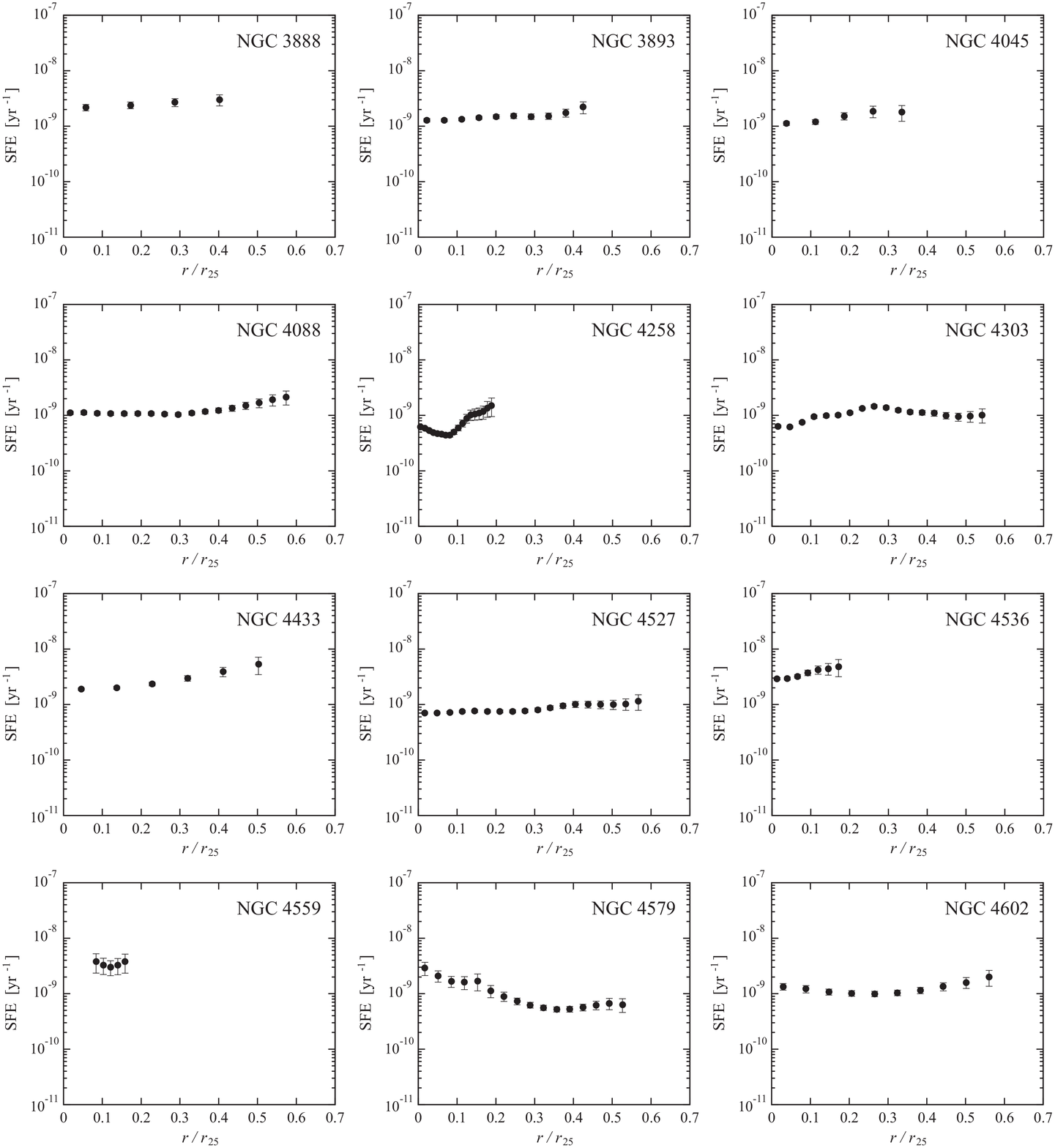}
  \end{center}
\caption{
(Continued)
}
\label{fig:fig4b}
\end{figure}

\setcounter{figure}{3}
\begin{figure}
  \begin{center}
    \includegraphics[width=17cm]{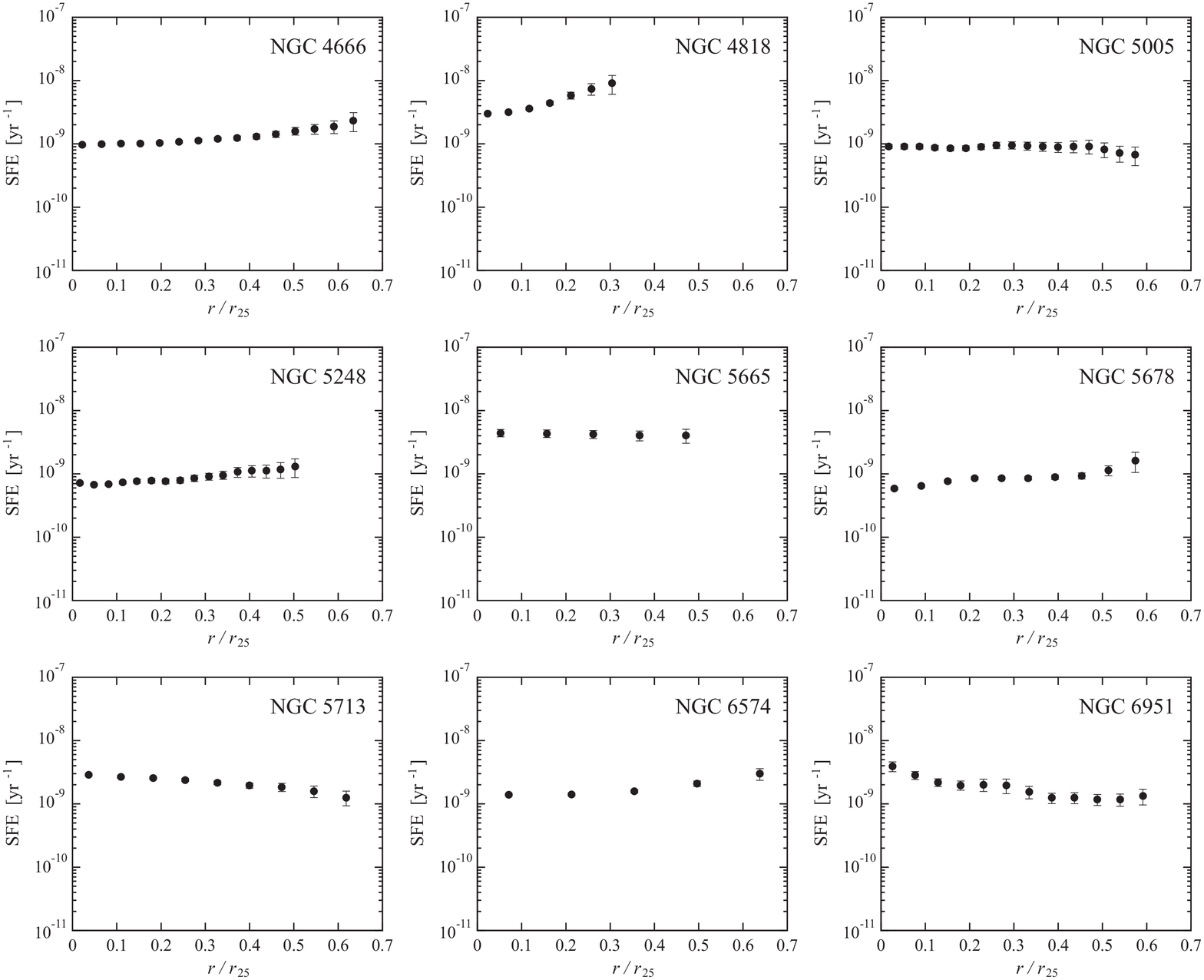}
  \end{center}
\caption{
(Continued)
}
\label{fig:fig4c}
\end{figure}

\begin{figure}
  \begin{center}
    \includegraphics[width=17cm]{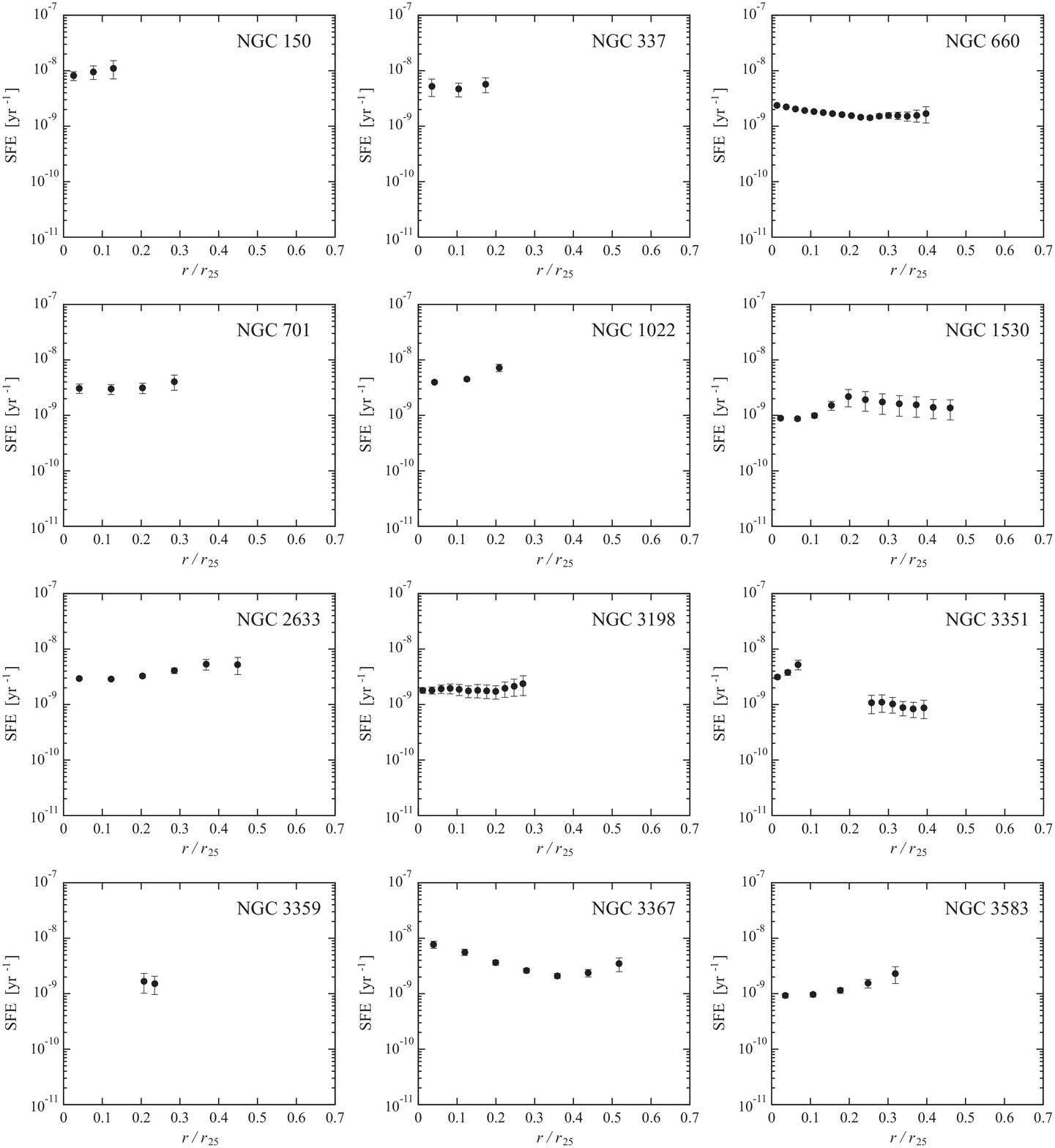}
  \end{center}
\caption{
SFE as a function of the galactocentric radius in the unit of $r/r_{25}$ for individual SB galaxies.
}
\label{fig:fig5a}
\end{figure}

\setcounter{figure}{4}
\begin{figure}
  \begin{center}
    \includegraphics[width=17cm]{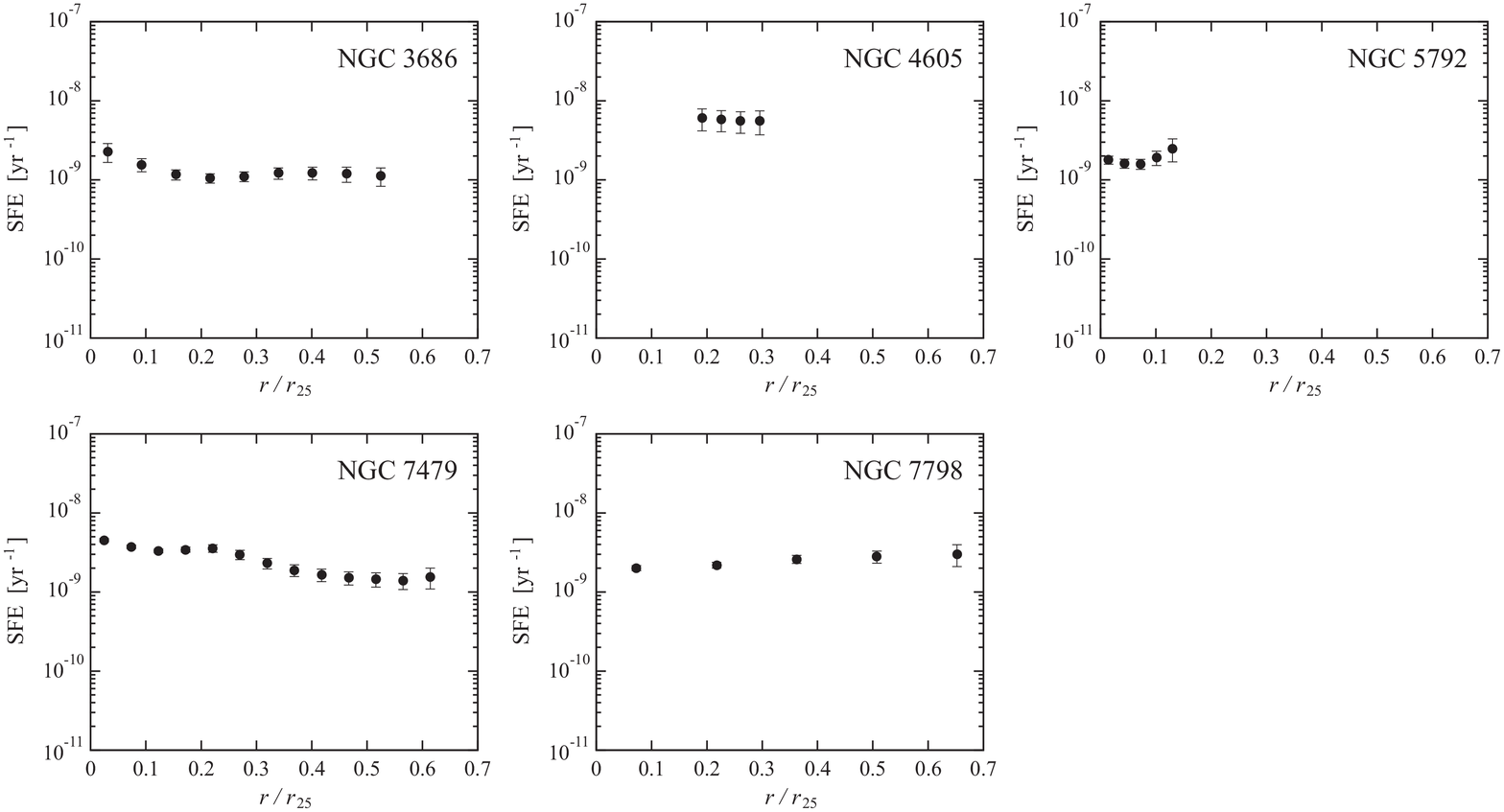}
  \end{center}
\caption{
(Continued)
}
\label{fig:fig5b}
\end{figure}

\begin{figure}
  \begin{center}
    \includegraphics[width=17cm]{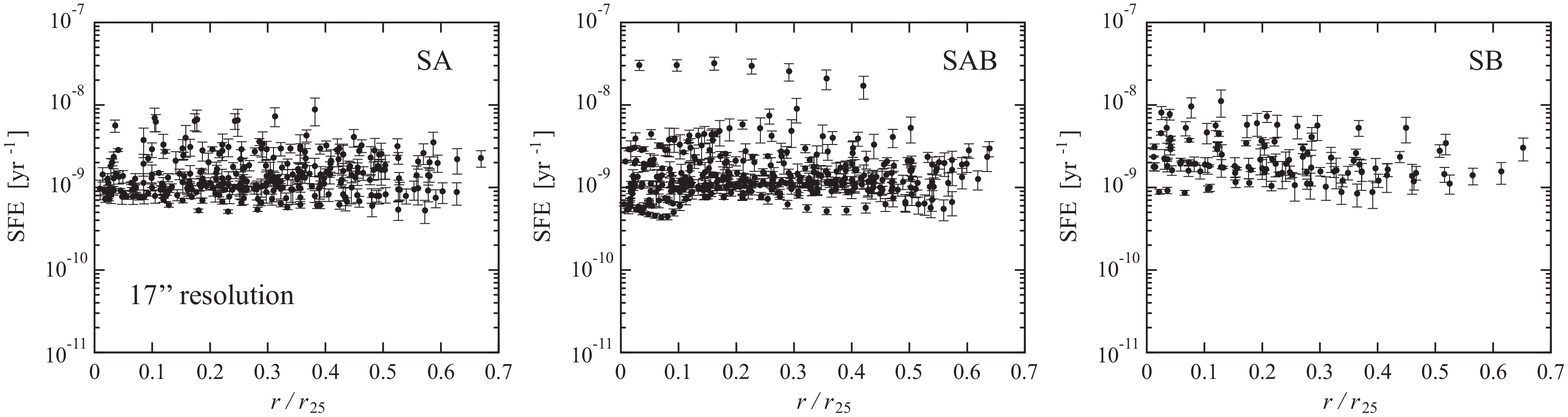}
  \end{center}
\caption{
Compiled radial profiles in SFE at a common angular resolution of 17'' for SA, SAB, and SB galaxies.
}
\label{fig:fig6}
\end{figure}

\begin{figure}
  \begin{center}
    \includegraphics[width=17cm]{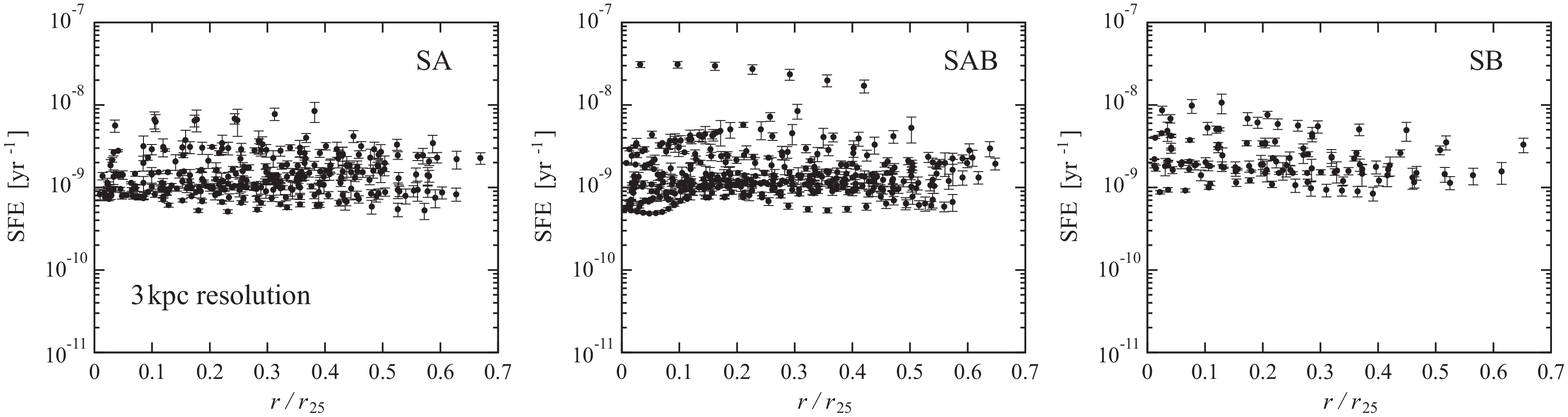}
  \end{center}
\caption{
Compiled radial profiles in SFE at a common spatial resolution of 3 kpc for SA, SAB, and SB galaxies.
}
\label{fig:fig7}
\end{figure}

\begin{figure}
  \begin{center}
    \includegraphics[width=8cm]{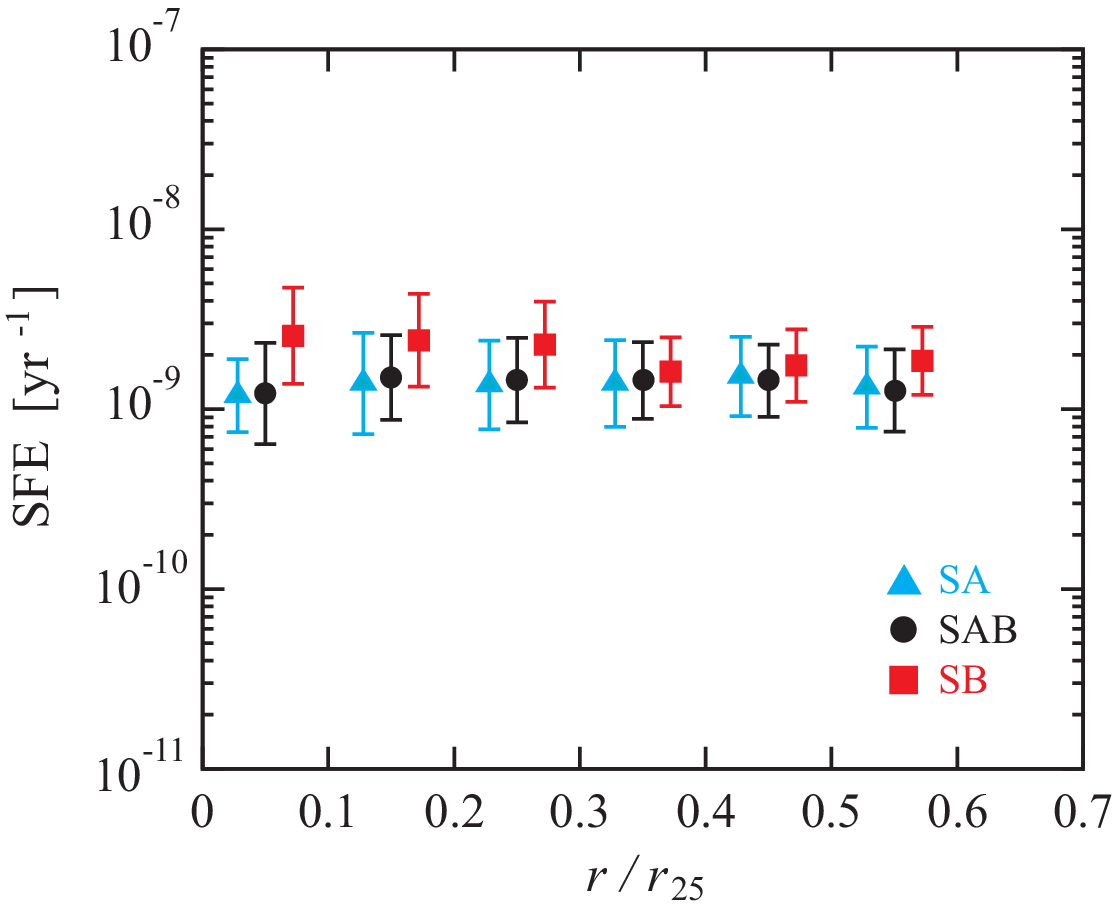}
  \end{center}
\caption{
Same as figure~6, but SFE is averaged over a 0.1 $r/r_{25}$ wide.
Blue triangles, black circles, and red squares indicate SA, SAB, and SB galaxies, respectively.
In order to make the difference in SFE among galaxy types more visible,
data points for SA and SB galaxies in each bin are offset to the left and the right side, respectively.
The error bars indicate the scatter of data points shown in figure 6.
}
\label{fig:fig8}
\end{figure}

\begin{figure}
  \begin{center}
    \includegraphics[width=8cm]{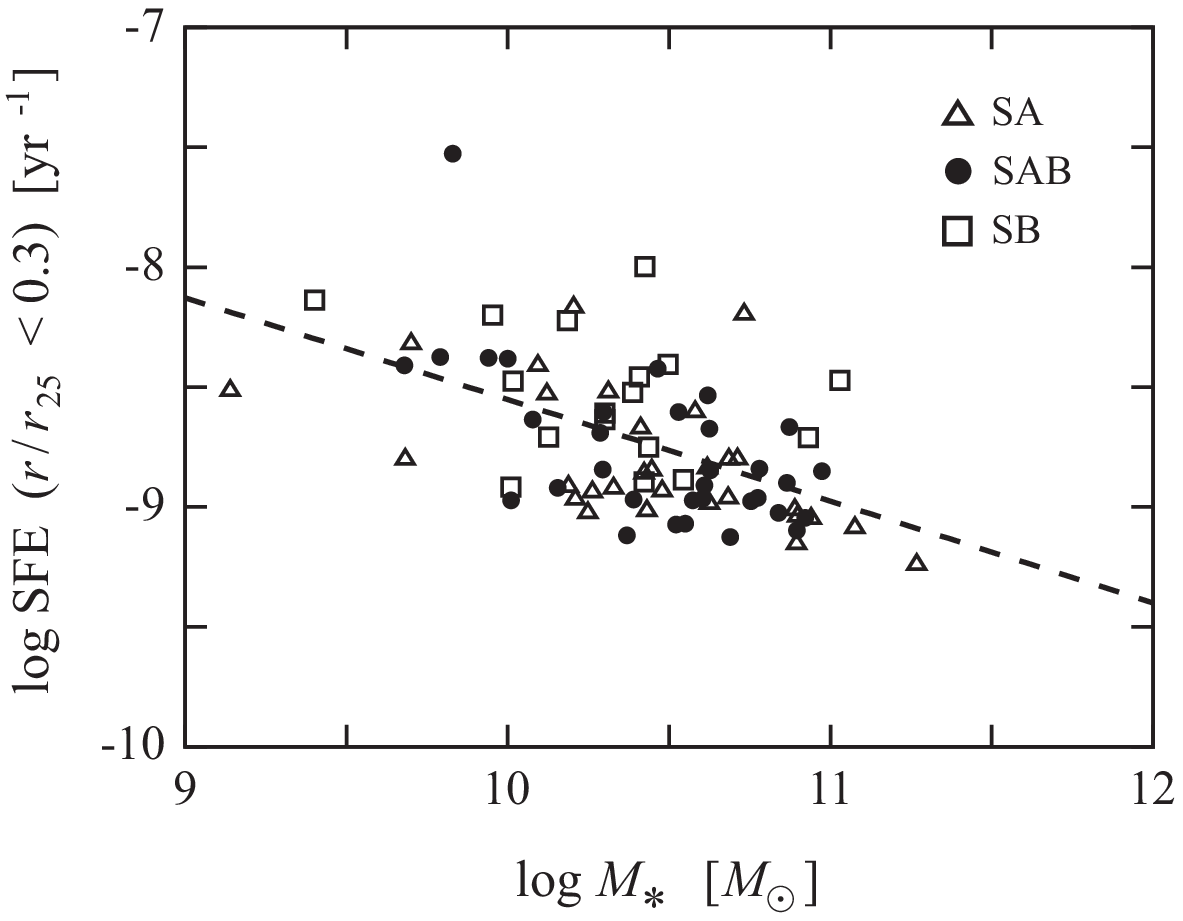}
  \end{center}
\caption{
Comparison between SFE within the inner radii ($r/r_{25} < 0.3$) and $M_{\ast}$ in our sample.
Open triangles, filled circles, and open squares indicate SA, SAB, and SB galaxies, respectively.
The dashed line indicates the ordinary least square power-law fit for all the data points, which is expressed as ${\rm log \,\, SFE_{fit}} = -0.424 \times ({\rm log} \,\, M_{\ast}) - 4.31$.
}
\label{fig:fig9}
\end{figure}

\begin{figure}
  \begin{center}
    \includegraphics[width=17cm]{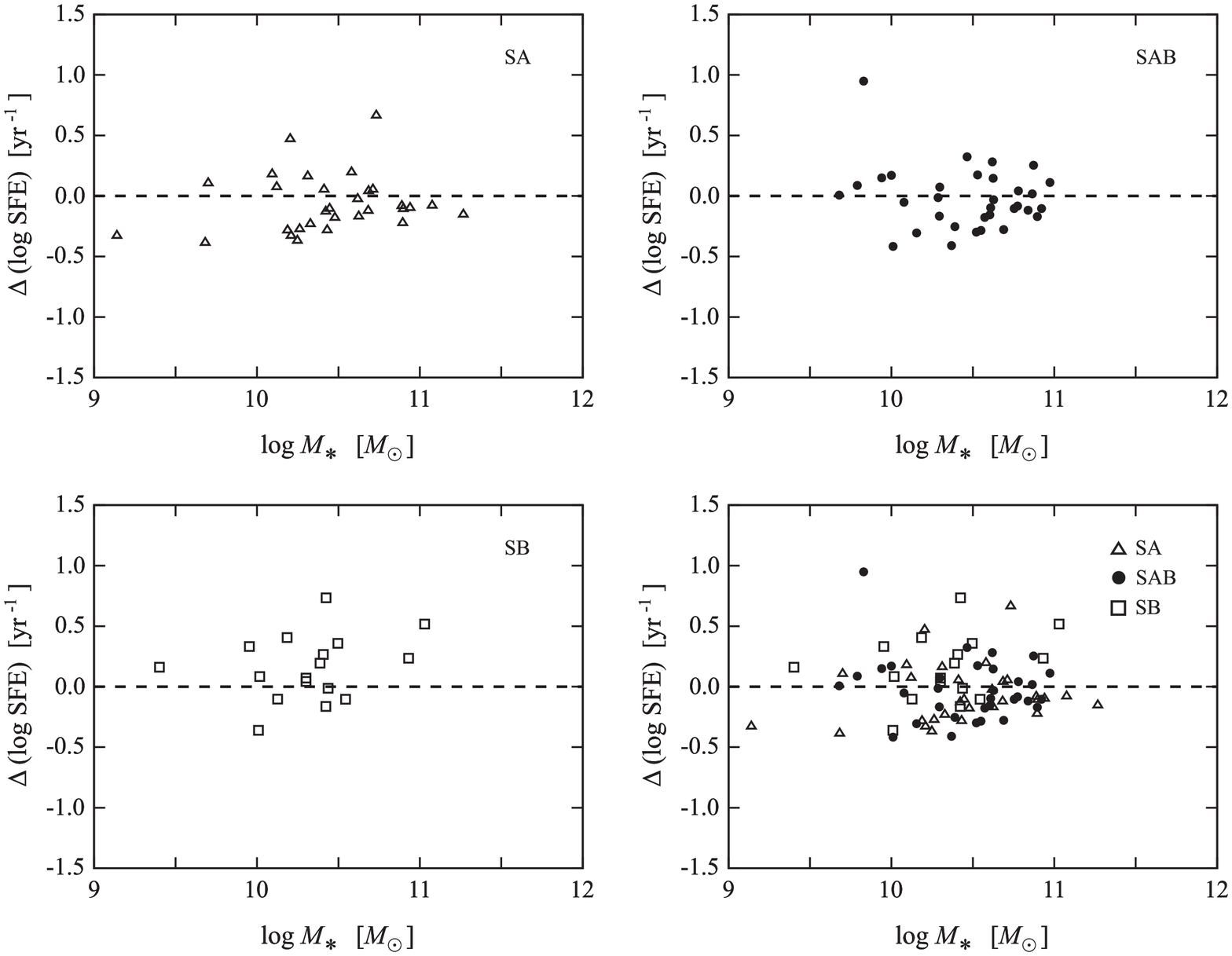}
  \end{center}
\caption{
Comparison between $\Delta$(log SFE) and $M_{\ast}$.
The dashed line indicates $\Delta$(log SFE) = 0.
Most of galaxies are within the $\Delta$(log SFE) range of $\pm$ 0.5.
}
\label{fig:fig10}
\end{figure}

\begin{figure}
  \begin{center}
    \includegraphics[width=8cm]{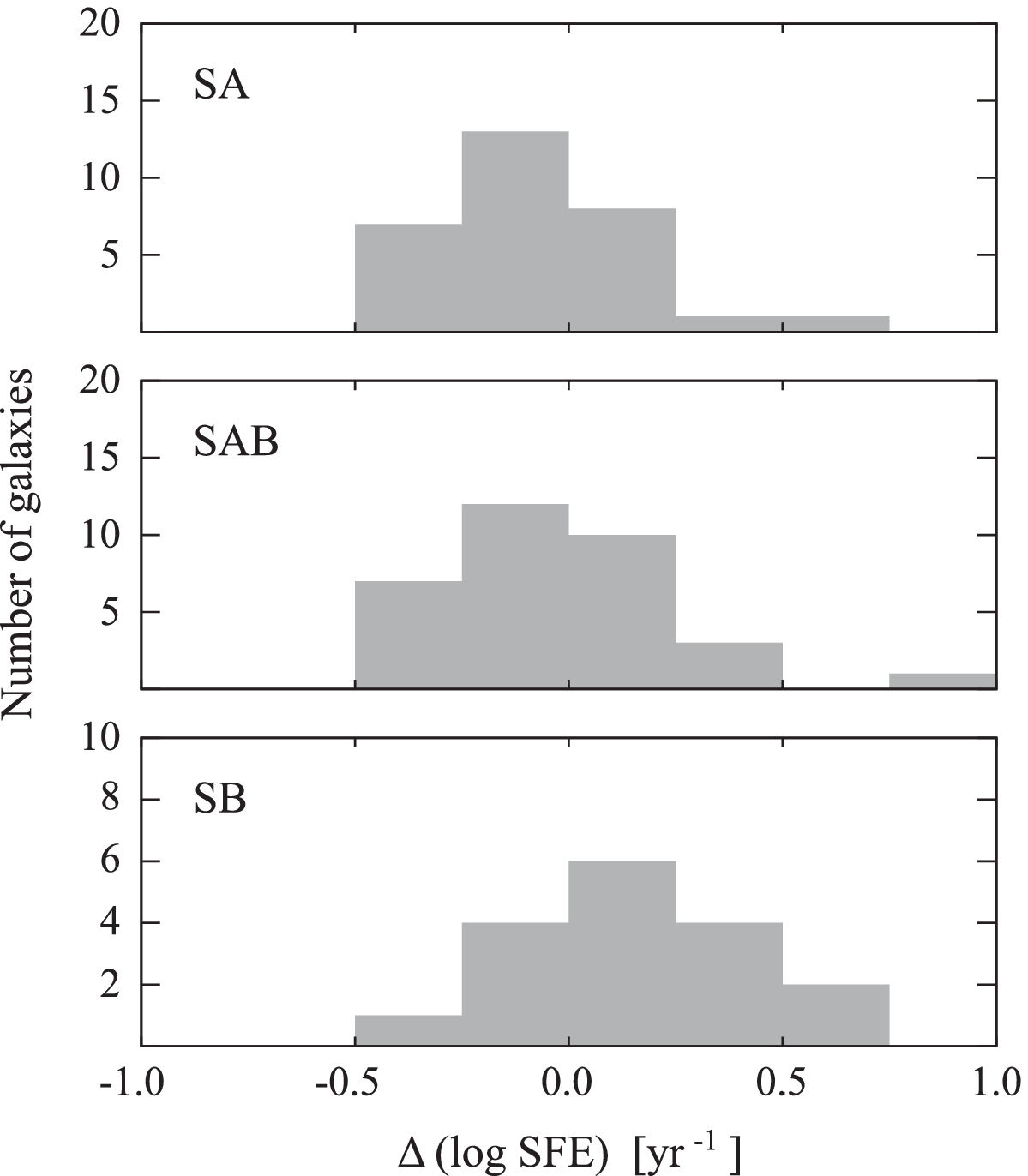}
  \end{center}
\caption{
Histograms of $\Delta$(log SFE) for SA, SAB, and SB galaxies.
}
\label{fig:fig11}
\end{figure}

\begin{figure}
  \begin{center}
    \includegraphics[width=17cm]{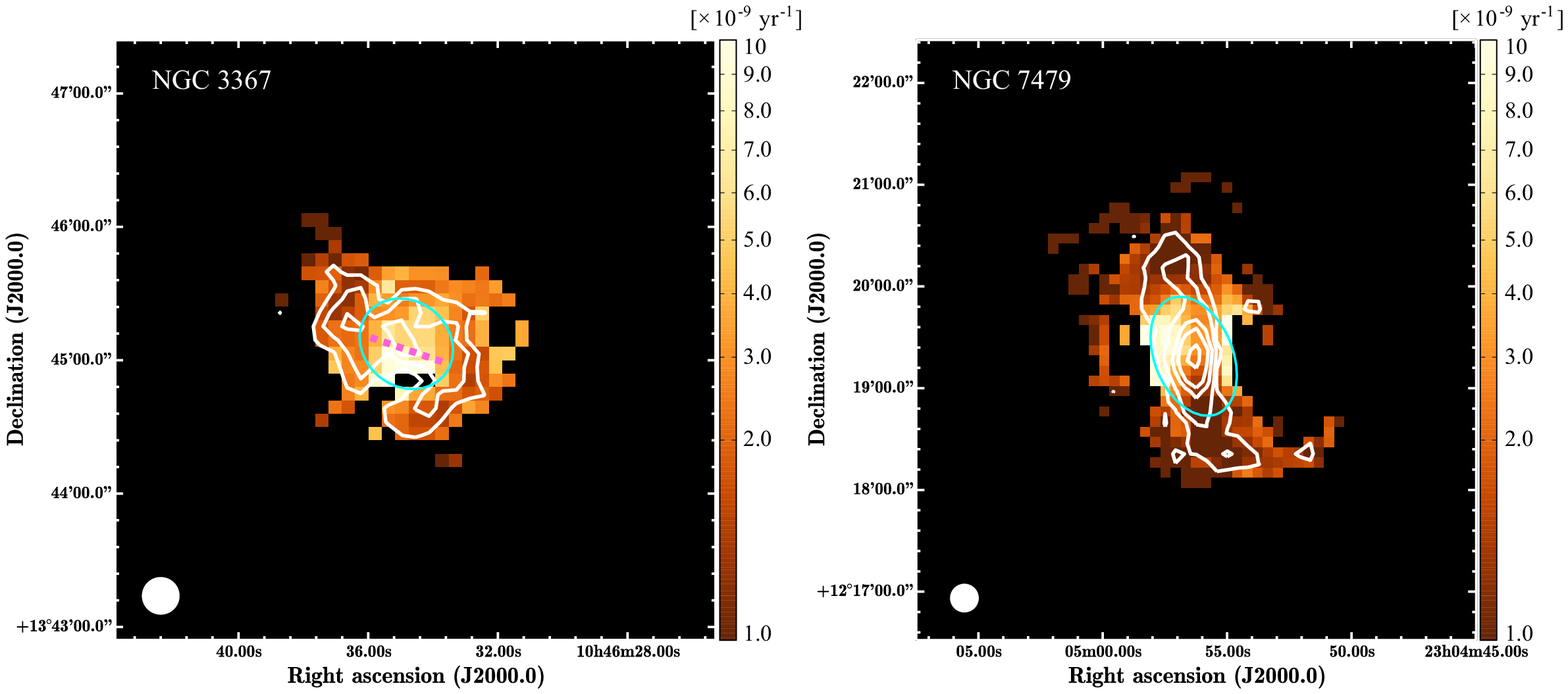}
  \end{center}
\caption{
Maps of SFE (color) superposed on the $^{12}$CO($J=1-0$) integrated intensity obtained by COMING (white contour) in NGC~3367 (left) and NGC~7479 (right).
The contour levels are 3, 6, and 10 K km s$^{-1}$ for NGC~3367 and 5, 10, 20, 30, and 50 K km s$^{-1}$ for NGC~7479, respectively.
The blue ellipse indicates $r/r_{25}$ = 0.3.
The white circle at bottom-left corner indicates an angular resolution of the maps, $17''$.
The dashed magenta line in NGC~3367 indicates the position of the bar.
}
\label{fig:fig12}
\end{figure}


\begin{table}
\begin{center}
Table~1.\hspace{4pt}Parameters of selected SA galaxies\\[1mm]

\begin{tabular}{lcccc}
\hline \hline \\[-6mm]
Galaxy   & $r_{25}$ & $i$         & P.A.         & log $M_{\ast}$ \\[-2mm]
         & (arcmin.)& ($^{\circ}$)& ($^{\circ}$) & ($M_{\odot}$)  \\
\hline \\[-6mm]
NGC 470   &  1.41  &  58.0  & 155.4  & 10.73  \\[-1mm]
NGC 628   &  5.24  &  7     & 20     & 10.25  \\[-1mm]
NGC 1084  &  1.62  &  57.2  & -141.6 & 10.58  \\[-1mm]
NGC 1482  &  1.23  &  55.2  & -61.5  & 10.31  \\[-1mm]
NGC 2742  &  1.51  &  59.9  & -93.5  & 10.42  \\[-1mm]
NGC 2748  &  1.51  &  72.8  & -138.8 & 10.12  \\[-1mm]
NGC 2775  &  2.14  &  35.4  & 163.5  & 10.71  \\[-1mm]
NGC 2841  &  4.07  &  73.7  & 152.6  & 10.90  \\[-1mm]
NGC 2967  &  1.51  &  16.5  & 64.0   & 10.21  \\[-1mm]
NGC 2976  &  2.95  &  64.5  & -25.5  & 9.14   \\[-1mm]
NGC 3147  &  1.95  &  35.2  & 142.79 & 11.27  \\[-1mm]
NGC 3169  &  2.19  &  39.0  & -123.7 & 10.89  \\[-1mm]
NGC 3338  &  2.95  &  60.9  & 97.1   & 10.48  \\[-1mm]
NGC 3370  &  1.58  &  55.1  & -38.1  & 10.09  \\[-1mm]
NGC 3655  &  0.78  &  23.5  & -100.3 & 10.68  \\[-1mm]
NGC 3672  &  2.09  &  67.2  & 7.8    & 10.63  \\[-1mm]
NGC 3675  &  2.95  &  67.8  & 176    & 10.89  \\[-1mm]
NGC 3810  &  2.14  &  42.2  & -154.3 & 10.33  \\[-1mm]
NGC 3813  &  1.12  &  68.2  & 83.1   & 10.26  \\[-1mm]
NGC 3938  &  2.69  &  20.9  & -154.0 & 10.43  \\[-1mm]
NGC 3949  &  1.44  &  52.9  & -58.2  & 10.20  \\[-1mm]
NGC 4030  &  2.09  &  39.0  & 29.6   & 11.08  \\[-1mm]
NGC 4041  &  1.35  &  23.4  & -138.7 & 10.68  \\[-1mm]
NGC 4632  &  1.55  &  65.9  & 60.5   & 9.70   \\[-1mm]
NGC 4750  &  1.02  &  40.0  & -50    & 10.62  \\[-1mm]
NGC 5364  &  3.38  &  47.9  & -144.4 & 10.45  \\[-1mm]
NGC 5676  &  1.99  &  59.8  & -131.9 & 10.94  \\[-1mm]
NGC 6503  &  3.54  &  73.5  & -60.2  & 9.68   \\[-1mm]
NGC 7625  &  0.79  &  37.4  & -151.4 & 10.19  \\[-1mm]
NGC 7721  &  1.78  &  69.8  & -164.2 & 10.41  \\
\hline \\[-2mm]
\end{tabular}\\
{\footnotesize
All the parameters are based on \citet{sorai2019}. They estimated $M_{\ast}$ based on the WISE 3.4 $\mu$m band infrared luminosity (\cite{wen2013}).}
\end{center}
\end{table}

\begin{table}
\begin{center}
Table~1. (continued)\hspace{4pt}Parameters of selected SAB galaxies\\[1mm]

\begin{tabular}{lcccc}
\hline \hline \\[-6mm]
Galaxy   & $r_{25}$ & $i$         & P.A.         & log $M_{\ast}$ \\[-2mm]
         & (arcmin.)& ($^{\circ}$)& ($^{\circ}$) & ($M_{\odot}$)  \\
\hline \\[-6mm]
NGC 157   &  2.09  &  48.0  & -136   & 10.16  \\[-1mm]
NGC 1087  &  1.86  &  50.5  & 1.4    & 10.00  \\[-1mm]
NGC 2268  &  1.62  &  58    & -112   & 10.62  \\[-1mm]
NGC 2276  &  1.41  &  48    & -113   & 10.87  \\[-1mm]
NGC 2715  &  2.45  &  67.8  & -159.1 & 10.01  \\[-1mm]
NGC 2903  &  6.30  &  67    & -155   & 10.61  \\[-1mm]
NGC 3166  &  2.40  &  55.7  & -100.4 & 10.78  \\[-1mm]
NGC 3310  &  1.55  &  56    & 150    & 9.83   \\[-1mm]
NGC 3344  &  3.54  &  27.0  & -37.3  & 10.08  \\[-1mm]
NGC 3368  &  3.80  &  57.5  & 169.0  & 10.52  \\[-1mm]
NGC 3437  &  1.26  &  65.9  & -61.5  & 10.29  \\[-1mm]
NGC 3627  &  4.56  &  52    & 176    & 10.60  \\[-1mm]
NGC 3888  &  0.87  &  41.8  & 121.2  & 10.53  \\[-1mm]
NGC 3893  &  2.24  &  30    & -13    & 10.29  \\[-1mm]
NGC 4045  &  1.35  &  48.4  & -92.1  & 10.63  \\[-1mm]
NGC 4088  &  2.88  &  68.9  & -126.8 & 10.39  \\[-1mm]
NGC 4258  &  9.31  &  72    & -29    & 10.55  \\[-1mm]
NGC 4303  &  3.23  &  27.0  & -36.4  & 10.77  \\[-1mm]
NGC 4433  &  1.10  &  64    & 5      & 10.63  \\[-1mm]
NGC 4527  &  3.09  &  70    & 69.5   & 10.69  \\[-1mm]
NGC 4536  &  3.80  &  64.2  & -54.5  & 10.46  \\[-1mm]
NGC 4559  &  5.36  &  63.1  & -36.8  & 9.68   \\[-1mm]
NGC 4579  &  2.95  &  41.7  & 92.1   & 10.84  \\[-1mm]
NGC 4602  &  1.70  &  67.9  & 100.7  & 10.75  \\[-1mm]
NGC 4666  &  2.29  &  70    & -135   & 10.57  \\[-1mm]
NGC 4818  &  2.14  &  67.2  & -175.5 & 9.94   \\[-1mm]
NGC 5005  &  2.88  &  66.7  & 67.0   & 10.92  \\[-1mm]
NGC 5248  &  3.09  &  38.6  & 103.9  & 10.37  \\[-1mm]
NGC 5665  &  0.96  &  51.7  & 154.7  & 9.79   \\[-1mm]
NGC 5678  &  1.66  &  56.9  & -177.5 & 10.89  \\[-1mm]
NGC 5713  &  1.38  &  33    & -157   & 10.30  \\[-1mm]
NGC 6574  &  0.71  &  45    & 165    & 10.97  \\[-1mm]
NGC 6951  &  1.95  &  30    & 135    & 10.87  \\
\hline \\[-2mm]
\end{tabular}\\
\end{center}
\end{table}

\begin{table}
\begin{center}
Table~1. (continued)\hspace{4pt}Parameters of selected SB galaxies\\[1mm]

\begin{tabular}{lcccc}
\hline \hline \\[-6mm]
Galaxy   & $r_{25}$ & $i$         & P.A.         & log $M_{\ast}$ \\[-2mm]
         & (arcmin.)& ($^{\circ}$)& ($^{\circ}$) & ($M_{\odot}$)  \\
\hline \\[-6mm]
NGC 150   &  1.95  &  59.3  & 108.9  & 10.42  \\[-1mm]
NGC 337   &  1.44  &  44.5  & 119.6  & 9.95   \\[-1mm]
NGC 660   &  4.16  &  72    & -138.5 & 10.44  \\[-1mm]
NGC 701   &  1.23  &  58.6  & 44.7   & 10.02  \\[-1mm]
NGC 1022  &  1.20  &  30    & 85     & 10.18  \\[-1mm]
NGC 1530  &  2.29  &  45    & -172   & 10.42  \\[-1mm]
NGC 2633  &  1.23  &  50.1  & -176.3 & 10.41  \\[-1mm]
NGC 3198  &  4.26  &  71.5  & -145.0 & 10.13  \\[-1mm]
NGC 3351  &  3.71  &  41    & -168   & 10.39  \\[-1mm]
NGC 3359  &  3.62  &  51    & -8     & 10.30  \\[-1mm]
NGC 3367  &  1.26  &  30    & 51     & 10.50  \\[-1mm]
NGC 3583  &  1.41  &  38.6  & 131.3  & 10.54  \\[-1mm]
NGC 3686  &  1.62  &  35.2  & 19.5   & 10.01  \\[-1mm]
NGC 4605  &  2.88  &  69    & -67    & 9.40   \\[-1mm]
NGC 5792  &  3.46  &  64    & -98.5  & 10.93  \\[-1mm]
NGC 7479  &  2.04  &  51    & -158   & 11.03  \\[-1mm]
NGC 7798  &  0.69  &  31.9  & 70.5   & 10.30  \\
\hline \\[-2mm]
\end{tabular}\\
\end{center}
\end{table}


\end{document}